\documentclass[11pt, a4paper]{article}

\pdfoutput=1
\usepackage{jheppub}
\usepackage[utf8]{inputenc}
\usepackage{epsfig}
\usepackage{graphicx}
\usepackage[multiple]{footmisc}
\usepackage{amssymb,amsmath}
\usepackage{fancybox,framed,tikz}
\usetikzlibrary{decorations.pathmorphing,patterns}
\usepackage{dsfont}
\usepackage{mathtools}
\usepackage{braket}
\usepackage{slashed}
\usepackage{rotating}
\usepackage{bbold,amsfonts}
\usepackage{multirow}
\usepackage{verbatim}

\newcommand{\beq}{\begin{equation}}
\newcommand{\eeq}{\end{equation}}

\renewcommand{\a}{\alpha}
\renewcommand{\b}{\beta}

\renewcommand{\d}{\delta}
\newcommand{\pa}{\partial}
\newcommand{\g}{\gamma}
\newcommand{\G}{\Gamma}

\newcommand{\D}{\Delta}
\newcommand{\e}{\epsilon}

\renewcommand{\k}{\kappa}

\renewcommand{\l}{\lambda}
\renewcommand{\L}{\Lambda}

\newcommand{\s}{\sigma}

\renewcommand{\t}{\tau}

\renewcommand{\O}{\Omega}

\newcommand{\Tr}{\textup{Tr}}
\newcommand{\Str}{\textup{Str}}
\newcommand\To{\rule{0pt}{2.6ex}}       

\author[a]{Lorenzo Bianchi,}
\author[b]{Luca Griguolo,}
\author[c]{Michelangelo Preti,}
\author[d]{Domenico Seminara}

\affiliation[a]{Institut f\"ur Theoretische Physik,
Universit\"at Hamburg\\
Luruper Chaussee 149,
22761 Hamburg, Germany}

\affiliation[b]{Dipartimento di Fisica e Scienze della Terra, Universit\`a di Parma and INFN
Gruppo Collegato di Parma, Viale G.P. Usberti 7/A, 43100 Parma, Italy}

\affiliation[c]{DESY Hamburg, Theory Group, Notkestra\ss e 85, 22607 Hamburg, Germany}

\affiliation[d]{Dipartimento di Fisica, Universit\`a di Firenze and INFN Sezione di Firenze, via
G. Sansone 1, 50019 Sesto Fiorentino, Italy}

\emailAdd{lorenzo.bianchi@desy.de}
\emailAdd{luca.griguolo@pr.infn.it}
\emailAdd{michelangelo.preti@desy.de}
\emailAdd{seminara@fi.infn.it}

\abstract{We study operator insertions into 1/2 BPS Wilson loops in ${\cal N}=6$ ABJM theory and investigate their two-point correlators. In this framework, the energy emitted by a heavy  moving probe can be exactly obtained from some two-point coefficients of bosonic and fermionic insertions.  This allows us to confirm an early proposal \cite{Bianchi:2014laa} for computing the  Bremsstrahlung function in terms of certain supersymmetric circular Wilson loops, whose value might be accessible to localization techniques.   In the derivation of this result we also elucidate  the structure of protected multiplets in the relevant superconformal defect theory and perform an explicit two-loop calculation.}

\title{Wilson lines as superconformal defects in ABJM theory: a formula for the emitted radiation}
\begin{document}
\maketitle

\section{Introduction and results}
In the past few years much progress has been made on computing exactly non-trivial observables in superconformal gauge theories,
nicely interpolating between weak and strong coupling regimes. Particular attention has been devoted to the study of anomalous dimensions of local operators, quantities that at large $N$ can be powerfully calculated by means of integrability \cite{Gromov:2009bc,Bombardelli:2009ns,Arutyunov:2009ur}. On the other hand, localization techniques \cite{Pestun:2016zxk} have produced exact, finite $N$ results for partition functions on curved manifolds \cite{Pestun:2007rz,Kapustin:2009kz,Benini:2012ui,Doroud:2012xw,Kallen:2012va,Minahan:2015jta} and non-local supersymmetric operators as Wilson or 't Hooft loops \cite{Pestun:2007rz,Kapustin:2009kz,Giombi:2009ek,Gomis:2011pf}. A natural generalization of this line of  investigation is to incorporate anomalous dimensions into particular Wilson loop operators and to exploit the  above techniques in their evaluation. There are two basic options on the market: the presence of cusps and/or operator insertions into Wilson loops may produce  divergences in perturbation theory, implying the appearance of anomalous dimensions \cite{Polyakov:1980ca,Korchemsky:1987wg,Drukker:2006xg}. This strategy was indeed considered in ${\cal N}=4$ SYM introducing cusps and/or operator insertions into 1/2 BPS Wilson lines and circular Wilson loops: in particular a set of boundary TBA ansatz equations \cite{Correa:2012hh} that calculate their spectrum was derived, leading to a solution for the quark-antiquark potential in this theory \cite{Gromov:2016rrp}. Moreover, taking the small angle limit of a cusped Wilson loop, one can define the so-called Bremsstrahlung function $B(\lambda)$ that computes the energy radiated by a moving quark in the low energy regime \cite{Fiol:2012sg,Correa:2012at,Correa:2012hh}. 
This is a non-trivial function of the coupling and can be calculated exactly using localization \cite{Fiol:2012sg,Correa:2012at} (see also \cite{Fiol:2015spa} for the case of $\mathcal{N}$=2 supersymmetry) or solving the boundary TBA equations in the appropriate limit\footnote{The TBA equations actually refer to a more general system with a local operator inserted at the tip of the cusp. Notably also this configurations is accessible by localization \cite{Giombi:2012ep,Bonini:2015fng}.} \cite{Gromov:2012eu,Gromov:2013qga}. The comparison of these two results allows to determine the interpolating function $h(\lambda)$ which features all the integrability computations. Whereas for $\mathcal{N}=4$ SYM this function is trivial, for ABJ(M) theory weak \cite{Gaiotto:2008cg,Grignani:2008is,Nishioka:2008gz,Minahan:2009aq,Minahan:2009wg,Leoni:2010tb} and strong \cite{McLoughlin:2008he, Abbott:2010yb,LopezArcos:2012gb,Bianchi:2014ada} coupling results showed a non-trivial dependence on the coupling and a conjecture for the exact form of this function was recently put forward in \cite{Gromov:2014eha} (see also \cite{Cavaglia:2016ide} for the generalization to ABJ theory).

 Interestingly, the Bremsstrahlung function is also obtained by inserting into the straight 1/2 BPS Wilson line a suitable operator and computing its two-point function \cite{Drukker:2011za,Correa:2012at}: the relevant object is called displacement operator \cite{Correa:2012at} and generates the small deformations of the Wilson line (for a recent thorough discussion of its role in the contest of defect CFTs see \cite{Billo:2016cpy} and references therein). More generally, operator insertions are organized according to the symmetry preserved by the 1/2 BPS Wilson line \cite{Cooke:2017qgm}: the set of correlation functions obtained in this way defines a defect conformal field theory. Anomalous dimensions and structure constants are associated to operator insertions in the usual way but  further data come into play: they are  the coefficients of the two-point functions, that are well defined physical quantities since the operators are normalized in the theory without the defect\footnote{In general the spectrum of a conformal defect is not related to that of the original CFT and one may have defect operators that were not present in the theory without the defect. Nevertheless, Wilson loops are given by the holonomy of a connection built out of gauge and matter fields and possible defect operators were already present in the original CFT.}. Recently this defect conformal field theory was considered from the point of view of AdS/CFT correspondence \cite{Giombi:2017cqn} and concrete strong coupling computations were successfully compared with localization results.

We would like to extend the above investigations to the three-dimensional ${\cal N }= 6$ superconformal ABJM theories \cite{Aharony:2008ug} but a first difference with the four-dimensional case immediately arises: in ABJM models not only bosonic but also fermionic matter can be used to construct generalized (super-)connections whose holonomy generates supersymmetric loop operators. Supersymmetric Wilson lines can be obtained using a generalized gauge connection that includes couplings to bosonic matter only, preserving 1/6 of the original supersymmetries \cite{Drukker:2008zx,Chen:2008bp}, while adding local couplings to the fermions the operator is promoted to be 1/2 BPS \cite{Drukker:2009hy}. The latter is dual to the fundamental string on $AdS_4\times CP^3$. Cusped Wilson loops formed with 1/6 BPS rays or 1/2 BPS rays are actually different \cite{Correa:2014aga} and, consequently, different Bremsstrahlung functions can be defined and, hopefully, evaluated exactly. In particular, in \cite{Lewkowycz:2013laa} a formula for the exact Bremsstrahlung function of the 1/6 BPS cusp was proposed, based on the localization result for the 1/6 BPS circular Wilson loop \cite{Kapustin:2009kz,Marino:2009jd,Drukker:2010nc,Klemm:2012ii}, and an extension to the 1/2 BPS case was argued. An exact expression for the Bremsstrahlung function of 1/2 BPS quark configurations was instead conjectured in \cite{Bianchi:2014laa}: this proposal was suggested by the analogy with the ${\cal N} = 4$ SYM case \cite{Correa:2012at} and supported by an explicit two-loop computation consistent with the direct analysis of the cusp with 1/2 BPS rays \cite{Griguolo:2012iq}. It was based on relating the Bremsstrahlung function with the derivative of some fermionic Wilson loop on a sphere $S^2$ with respect to the latitude angle \cite{Bianchi:2014laa}. Recently, a non-trivial three-loop test of the above proposal has been performed by computing the Bremsstrahlung function associated to the 1/2 BPS cusp in ABJM theory \cite{Bianchi:2017svd}: the final result precisely reproduces the formula appeared in \cite{Bianchi:2014laa} including color subleading corrections.

In this paper we take a different approach to the study of the Bremsstrahlung function for the 1/2 BPS cusp in ABJM theory, exploiting its definition in terms of two-point correlators inserted into the Wilson line. More generally, we initiate the investigation of the defect conformal field theory associated to the 1/2 BPS straight line in ${\cal N}=6$ superconformal Chern-Simons theory, very much in the same spirit of \cite{Correa:2012at,Cooke:2017qgm}. As already done in \cite{Drukker:2011za,Griguolo:2012iq}, we consider a cusped Wilson line depending on two parameters: the geometric euclidean angle $\varphi$ between the two 1/2 BPS lines defining the cusp, and an internal angle $\theta$ describing the change in the orientation of the couplings to matter between the two rays \cite{Drukker:2011za,Griguolo:2012iq}. At $\varphi^2 = \theta^2$ the cusped Wilson loop is BPS and its anomalous dimension vanishes. For small angles, the expansion of the cusp anomalous dimension around the BPS point reads
\begin{equation}
\Gamma_{\rm cusp}(\lambda,\theta,\varphi)\simeq(\theta^2-\varphi^2) B(\lambda)
\end{equation}
where $B(\lambda)$ is the Bremsstrahlung function, a non-trivial function of the coupling constant of the theory. From this equation we can read off the Bremsstrahlung function equivalently from the $\theta$ or the $\varphi$ expansions of $\Gamma_{\rm cusp}$, setting the other angle to zero. Working only with the internal angle $\theta$, we show that the Bremsstrahlung function can be extracted from the (traced) correlation function of two super-matrix operators inserted into the line. The computation is in turn reduced, using some symmetry consideration, to the evaluation of a bosonic and a fermionic two-point functions. The latter's kinematic part is fully determined by conformal symmetry and the entire dynamical content is encoded into two coefficients, $c_s$ and $c_f$. Therefore the Bremsstrahlung function is expressed for any coupling as a linear combination of those. 
The very same combination can be also obtained deforming the 1/2 BPS circular Wilson loop to the 1/6 BPS latitude and performing a suitable derivative with respect to the deformation parameter $\nu$. 
These results allow to justify the expression proposed in \cite{Bianchi:2014laa} 
\begin{equation}
 B(\l,N)=\frac{1}{4\pi^2} \left.\frac{\pa}{\pa \nu} \log\braket{\mathcal{W}_{\nu}}\right|_{\nu=1}
\end{equation}
and further confirm the possibility of an exact calculation by localization. 

A crucial step in our derivation is the non-renormalization properties of some operators inserted into the Wilson line: to prove them and to connect our procedure with the definition in terms of the displacement operator we carefully analyze the symmetry structure of the defect conformal field theory associated to the 1/2 BPS Wilson line and its representation theory. We find that the theory preserves $SU(1,1|3)$ whose bosonic subgroup is $SU(1,1)\times SU(3)\times U(1)_M$.  Defect operators (local operators of the full ABJM theory inserted along the line) are therefore characterized by a set of four quantum numbers $(\Delta,m,j_1,j_2)$ associated to the 4 Cartan generators of the bosonic subalgebra. We study the structure of short and long multiplets representing this subalgebra and we identify those associated to the defect operators relevant for our case. In the same supermultiplets we find some of the components of the displacement operator which we express as a super-matrix with operatorial entries. Since the scaling dimension of the displacement operator is fixed by a Ward identity all the components of its supermultiplet are protected, including those of interest for us.

As a check of our results we perform a concrete two-loop computation. We evaluate the Bremsstrahlung function using its relation to the two-point defect correlation functions of bosonic and fermionic operators and we compare it to the two-loop result of \cite{Griguolo:2012iq,Bianchi:2014laa}.

The plan of the paper is the following: in Section 2 we briefly recall the structure of the 1/2 BPS Wilson loops in ABJM theory, the construction of the generalized cusp and of deformed circular loop on $S^2$ in relation with the Bremsstrahlung function. In Section 3 we define the relevant defect correlation functions and describe the symmetry structure of the defect conformal field theory. In Section 4 we derive the expression of the Bremsstrahlung function in terms of defect correlators and explain the relation with a suitable derivative of the deformed circular Wilson loop. The super-displacement operator is instead studied in Section 5, where also the structure of its super-multiplet is discussed. Section 6 is devoted to the perturbative checks. Appendix A contains our conventions while in Appendix B we recall the $osp(6|4)$ algebra. Appendix C is quite important being devoted to the $SU(1,1|3)$ subalgebra that is the symmetry of our defect correlators: we discuss its representations and the displacement multiplets. In Appendix D we write down for completeness the supersymmetry transformations of the theory.

\section{Wilson loops in ABJM and Bremsstrahlung function}\label{WLABJM}
The task of finding supersymmetry preserving line operators in ABJM theory is notably more intricate than in the four-dimensional relative $\mathcal{N}=4$ SYM. The first proposal for a supersymmetric Wilson loop was put forward in \cite{Drukker:2008zx,Chen:2008bp,Rey:2008bh} as a natural generalization of the four-dimensional Wilson-Maldacena loop \cite{Maldacena:1998im}. However, in three dimensions, such Wilson loop turns out to be 1/6 BPS and in order to obtain a 1/2 BPS object one needs to introduce couplings with fermions and consider the holonomy of a superconnection of the $U(N|N)$ supergroup \cite{Drukker:2009hy,Lee:2010hk}. A generalization of this construction for arbitrary contour was given in \cite{Cardinali:2012ru} where the Wilson loop was expressed as
\begin{equation}\label{WL}
 \mathcal{W}=\Str\left[P\exp\left(-i\oint d\tau \mathcal{L}(\t) \right) \mathcal{T}\right]
\end{equation}
with a superconnection $\mathcal{L}(\t)$
\begin{equation}\label{superconnection}
\mathcal{L}=\begin{pmatrix} A_\mu \dot{x}^\mu-\frac{2\pi i}{k} |\dot x| {M_J}^I C_I \bar C^J & -i\sqrt{\frac{2\pi}{k}}|\dot{x}|\eta_I \bar \psi^I\\
             -i\sqrt{\frac{2\pi}{k}}|\dot{x}| \psi_I \bar \eta^I & \hat A_\mu \dot x^\mu-\frac{2\pi i}{k} |\dot x| {\hat{M}_J}^I  \bar C^J C_I
            \end{pmatrix}
\end{equation}
Here the contour of the loop is parametrized by $x^{\mu}(\tau)$ and the quantities ${M_J}^I(\tau)$, ${\hat{M}_J}^I$, $\eta_I(\tau)$ and $\bar \eta^I(\tau)$ are local couplings, whose form is  determined in terms of the contour $x^{\mu}(\tau)$ by the requirement of preserving some of the supercharges. The key idea of \cite{Drukker:2009hy,Lee:2010hk,Cardinali:2012ru} is to relax the condition $\d_\text{susy}\mathcal{L}=0$ and replace it with the weaker requirement
\begin{equation}\label{susyvarL}
\d_\text{susy}\mathcal{L}=\mathcal{D}_{\tau} \mathcal{G}=\pa_{\tau} \mathcal{G}+i[\mathcal{L},\mathcal{G}] 
\end{equation}
where $\mathcal{G}$ is a $u(N|N)$ supermatrix. This implies a vanishing variation for the (super)traced Wilson loop, provided the correct periodicity of $\mathcal{G}$. The twist supermatrix $\mathcal{T}$ in \eqref{WL} is introduced with the precise aim of closing the loop after a supersymmetry transformation and its defining equation is
\begin{equation}
 \mathcal{T}\mathcal{G}(\tau_0)=\mathcal{G}(0)\mathcal{T}
\end{equation}
where $\tau_0$ is the period of the loop. In the following we shall be interested in two particular configurations: the generalized cusp and the 1/6 BPS latitude Wilson loop. The latter is a two-parameter deformation of the 1/2 BPS circular Wilson loop running around the sphere \cite{Bianchi:2014laa}. We will comment more on the contour of the loop in section \ref{circular}. Let us introduce the two Wilson loop configurations separately. A summary of notations and conventions is given in appendix \ref{conventions}.

\subsection{The generalized cusp}
\begin{figure}[htbp]
\begin{center}
\begin{tikzpicture}[scale=0.55]
  \draw[double,thick,blue]  (-2.6,0)--(0,0);
  \draw[->,double,thick,blue]  (-5,0)--(-2.5,0);
  \draw[double,thick,blue]  (2.48,1.24)--(5,2.5);
  \draw[->,double,thick,blue]  (0,0)--(2.5,1.25);
  \draw (0,0)--(5,0);
  \draw[->] (3.5,0) arc (0:26:3.5);
  \node[above] at (2,0) {$\varphi$};
  \draw (7,-2)--(7,5);
    \draw[double,thick,blue] (7.5,0.4)--(7.5,4.5);
  \draw[->,double,thick,blue] (7.5,-3.3)--(7.5,0.5);
      \draw[double,thick,blue] (12,0.2)--(12,4.3);
  \draw[-<,double,thick,blue] (12,-3.5)--(12,0.4);
  \draw[dashed] (7.5,-2.8)--(12.5,-1.2);
  \draw[dashed] (10,-2)--(12,-3.1);
  \draw (13,-2)--(13,5);
  \draw (10,5) ellipse (3 and 1.5);
  \draw (10,-2) ellipse (3 and 1.5);
  \node at (11.2,-2.1) {$\varphi$};
  \draw[->] (11.5,-1.5) arc (30:-45:1);
  \node at (11.2,-2.1) {$\varphi$};
 \end{tikzpicture}
 \caption{The cusp setting on the plane and on the cylinder. The two configurations are mapped to each other by a conformal transformation and this relates the cusp anomalous dimension to the quark-antiquark potential for any conformal gauge theory.} \label{fig:cusp}
 \end{center}
\end{figure}
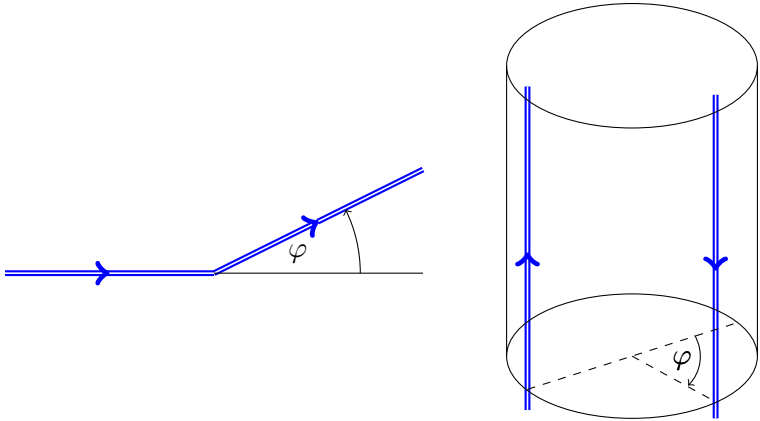

Let us start by deforming the straight Wilson line by a generalized cusp. This configuration was first introduced in \cite{Drukker:1999zq,Drukker:2011za} for $\mathcal{N}=4$ SYM and then adapted to ABJM theory in \cite{Griguolo:2012iq}. As pictured in figure \ref{fig:cusp}, a cusp on the plane $\mathbb{R}^3$ can be conformally mapped to a pair of anti-parallel lines on $S^2\times \mathbb{R}$. 

In the generalized cusp setting we introduce an additional angle $\theta$, which we take to be non-vanshing only on one branch of the cusp (or equivalently on one of the two lines on the cylinder). This additional angle is purely internal and does not affect the physical contour of the line. In particular for the setting in figure \ref{fig:cusp} the couplings in the superconnection \eqref{superconnection}, for the second branch of the cusp, read
\begin{equation}\label{boslinecoupling}
 {M_I}^J={\hat{M}_I}{}^J=\begin{pmatrix}
                    -\cos\theta 	&\sin\theta	&0	&0\\
                    \sin\theta	&\cos\theta		&0	&0\\
                    0		&0				&1	&0\\
                    0		&0				&0	&1\\
                   \end{pmatrix} \, , \qquad
\mathcal{T}= \begin{pmatrix}
              \mathbb{1}_N 	& 0\\
              0						&-\mathbb{1}_N 
             \end{pmatrix} 
\end{equation}
\begin{equation}\label{ferlinecoupling}
\eta_I^\a=
                   \begin{pmatrix}
                    \cos\frac{\theta}{2}\\
                    -\sin\frac{\theta}{2}\\
                    0\\
                    0
                   \end{pmatrix}_I
                   \sqrt{2}\eta^+ \, ,\quad 
\bar\eta^I_\a=i
                   \begin{pmatrix}
                    \cos\frac{\theta}{2}&
                     -\sin\frac{\theta}{2}&
                    0&
                    0
                   \end{pmatrix}^I
                   \sqrt{2}\bar\eta_+
\end{equation}
where $\eta^+=\bar\eta_+^T=\frac{1}{\sqrt{2}}\begin{pmatrix}
                    1 & 1
                    \end{pmatrix}$. 
                    
In general, deforming a straight line by a cusp  does not preserve any of the supersymmetries of the original setting, but it is still very interesting from a physical point of view. The expectation value of the cusped Wilson loop is logarithmically divergent and can be parametrized as
\begin{equation}\label{Gammacusp}
 \log \braket{\mathcal{W}}\sim-\G_{\rm cusp}(\theta,\varphi,\l,N) \log\frac{L}{\e}+\mbox{finite}
\end{equation}
The coefficient of the logarithm is the celebrated cusp anomalous dimension, controlling IR divergences for scattering amplitudes of massive colored particles.  Here  $L$ is identified with the infrared cut-off while $\epsilon$ with the ultraviolet one.
Recently, $\Gamma_{cusp}$ was studied in depth at weak and strong coupling. Its value is known up to two loops via perturbation theory \cite{Griguolo:2012iq} and exactly in the scaling limit where only ladder diagrams contribute ($\lambda\rightarrow 0$, $i\theta\rightarrow\infty$ and $\lambda e^{i\theta/2}=\text{const}$) \cite{Bonini:2016fnc}. Using the HQET formalism, $\Gamma_{cusp}$ was computed up to three-loop in the $\varphi=0$ case \cite{Bianchi:2017svd}. At strong coupling, it was studied up to next to leading order in \cite{Forini:2012bb,Correa:2014aga}.
Furthermore, the plane to cylinder mapping relates the cusp to antiparallel lines, whose  expectation value yields the generalized potential between a quark and an anti-quark sitting on $S^2$ at an angle $\pi-\varphi$
\begin{equation}\label{potential}
  \log \braket{\mathcal{W}}\sim-V(\theta,\varphi,\l,N)\, T
\end{equation}
where $T$, the length of the lines, is the IR cut-off in this setting. The precise change of coordinates relating the two cut-offs is made explicit, together with a precise list of conventions, in appendix \ref{conventions}. 

There are also some interesting limits of the generalized cusp anomalous dimension. Let us review some of them:
\begin{itemize}
\item First we consider the (almost trivial) limit of vanishing angle, in which case we recover the 1/2 BPS straight line configuration. In section \ref{symmetry} and appendix \ref{subalgebra} we carry out a thorough analysis of the supergroup preserved by this configuration.
\item By analytic continuation one can investigate the cusp anomalous dimension for imaginary values of the angle $\varphi$ relating it, for infinite purely imaginary $\varphi$, to the light-like cusp anomalous dimension, whose value has been famously computed  exactly using integrability \cite{Beisert:2006ez,Gromov:2008qe}.
 \item When the physical angle $\varphi$ equals the internal angle $\theta$ one finds another BPS configuration and the expectation value of the line is no longer divergent\cite{Drukker:2011za,Griguolo:2012iq}
 \begin{equation}\label{BPScond}
 \G_{cusp}(\varphi,\varphi,\l,N)=0
 \end{equation}
\item As mentioned in the introduction, expanding the cusp anomalous dimension for small angles, imposing parity and using \eqref{BPScond}, one finds
\begin{equation}\label{smallangle}
 \G_{cusp}(\theta,\varphi,\l,N) \sim B(\l,N) (\theta^2-\varphi^2)
\end{equation}
where $B(\l,N)$ is the Bremsstrahlung function, so called because it determines the energy emitted by a moving heavy probe \cite{Correa:2012at}. One of the goal of this paper is to relate this function to the expectation value of some circular Wilson loop.
\end{itemize}

\subsection{Circular Wilson loops}\label{circular}
It is a well-known fact that, despite being related by a conformal transformation, the expectation value of a circular and a straight line 1/2 BPS Wilson loops are not the same \cite{Erickson:2000af,Drukker:2000rr,Pestun:2007rz}. In particular for ABJM theory, a matrix model for the 1/6 BPS case was derived in \cite{Kapustin:2009kz} and solved in \cite{Marino:2009jd,Drukker:2010nc}. The 1/2 BPS circular Wilson loop is also known exactly thanks to its cohomological equivalence with a linear combination of the 1/6 BPS ones \cite{Drukker:2008zx}. The question of deforming the maximally supersymmetric 1/2 BPS configuration preserving some supersymmetry was thoroughly addressed in \cite{Cardinali:2012ru,Bianchi:2014laa} where, for instance, a two-parameter 1/6 BPS deformation was written down. In that case the deformation for the circle was derived by moving the contour on the sphere $S^2$ from the equator to a latitude with angle $\theta_0$. In general this has consequences both on the integration contour and on the explicit expression of the superconnection, whose coupling can be written in terms of the contour. However let us stress that moving the physical contour from the equator to the latitude has no effect on the expectation value of the Wilson loop since it is equivalent to a conformal transformation. Therefore the only variation that is actually relevant for the expectation value of the loop is that in the superconnection. In the following this property will prove crucial for our arguments. 
The second parameter of the deformation is a purely internal angle $\alpha$ whose value range in the interval $[0,\frac{\pi}{4}]$ \cite{Cardinali:2012ru,Bianchi:2014laa}. Despite apparently dependent on two parameters, it turns out one can write down the superconnection \eqref{superconnection} in terms of a single parameter
\begin{equation}
 \nu=\sin 2\a \cos \theta_0
\end{equation}
with couplings given by 
\begin{equation}\label{latcoupling}
{M_I}^J=\hat M_I{}^J=\begin{pmatrix}
                    -\nu 	&e^{-i\tau} \sqrt{1-\nu^2}	&0	&0\\
                    e^{i\tau} \sqrt{1-\nu^2}	&\nu		&0	&0\\
                    0		&0				&1	&0\\
                    0		&0				&0	&1\\
                   \end{pmatrix} \, , \qquad
\mathcal{T}= \begin{pmatrix}
              e^{-\frac{i\pi\nu}{2}} \mathbb{1}_N 	& 0\\
              0						&e^{\frac{i\pi\nu}{2}} \mathbb{1}_N
             \end{pmatrix}
\end{equation}
\begin{align}
\eta_I^\a&=\textstyle{e^{\frac{i\nu \tau}{2}}}
                   \begin{pmatrix}
                    \textstyle{\sqrt{\tfrac{1+\nu}{2}}}\\
                    -\textstyle{\sqrt{\tfrac{1-\nu}{2}}} \textstyle{e^{i\tau}}\\
                    0\\
                    0
                   \end{pmatrix}_I
                   \begin{pmatrix}
                    1 & -ie^{-i\tau}
                    \end{pmatrix}^\a ,&
\bar\eta^I_\a&=\textstyle{ie^{\frac{-i\nu \tau}{2}}}
                   \begin{pmatrix}
                    \sqrt{\tfrac{1+\nu}{2}}&
                    -\sqrt{\tfrac{1-\nu}{2}} \textstyle{e^{-i\tau}}&
                    0&
                    0
                   \end{pmatrix}^I
                   \begin{pmatrix}
                    1 \\ ie^{i\tau}
                    \end{pmatrix}_\a\nonumber
\end{align}
Because of our previous argument on the independence of the expectation value on $\theta_0$ through the contour, we can safely conclude that the expectation value of the loop will depend only on $\nu$. In the following we will show that the derivative of this Wilson loop with respect to the parameter $\nu$ gives the Bremsstrahlung function.

\section{Symmetry considerations}\label{symmetry}
Before starting the derivation, let us make some considerations on the symmetries preserved by the Wilson line. We focus on the straight line case, but all our consideration can be extended, by conformal mapping, to the circular case. The 1/2 BPS Wilson line breaks the $OSP(6|4)$ symmetry down to a $SU(1,1|3)$ subgroup. Consequently the operators of the theory, in presence of the line, reorganize themselves into representations of $SU(1,1|3)$. 

The bosonic subgroup of $SU(1,1|3)$ is $SU(1,1)\times SU(3)\times U(1)_M$. The first factor is simply the conformal algebra in one dimension, generated by $\{P_1,K_1,D\}$. According to this symmetry we split the spacetime coordinates as 
\begin{equation}\label{coordsplit}
 x^{\mu}=(x^1,x^m)
\end{equation}
with $m=2,3$. Occasionally we will find convenient to use complex coordinates in the orthogonal directions
\begin{equation}
 z=\frac{x^2+i x^3}{\sqrt{2}} \qquad \bar z=\frac{x^2-i x^3}{\sqrt{2}}
\end{equation} 
Moreover, fermions will be expressed in a basis of eigenvalues of $\gamma_1=\sigma_1$ (see appendix \ref{conventions})
\begin{equation}
 \psi_+=\frac{1}{\sqrt2}(\psi_1+\psi_2) \qquad \psi_-=\frac{1}{\sqrt2}(\psi_1-\psi_2)
\end{equation}
with the rules $\psi^-=-\psi_+$ and $\psi^+=\psi_-$.

The $SU(3)$ subgroup of $SU(1,1|3)$ is the residual R-symmetry group generated by a subset ${R_{a}}^b$ ($a,b=1,2,3$) of the former $SU(4)$ generators ${{J_I}}^K$ as shown explicitly in appendix \ref{subalgebra}. Finally, the $U(1)_M$ factor is a recombination of the rotation around the line and a broken R-symmetry generator. This can be understood by the following argument. The fermionic couplings $\eta$ and $\bar\eta$ in \eqref{superconnection} break rotational symmetry in the orthogonal plane. Nevertheless for the straight-line case the couplings are particularly simple (as one can immediately observe by taking the limit $\theta\to 0$ in \eqref{boslinecoupling} and \eqref{ferlinecoupling}) and the only fermionic combination appearing in the superconnection \eqref{superconnection} (see also \eqref{straightlineconn} below) is $\psi^1_+$ together with its conjugate. This combination is an eigenstate under rotations around the line (generated by $M_{23}$ in the notation of appendix \ref{algebra}) and under the action of the broken R-symmetry generator ${J_1}^1$. It is not hard to check that the combination\footnote{The coefficient in the linear combination below are due to the normalization of the generators (see appendix \ref{algebra}).}  
\begin{equation}
M=3i M_{23}-2 {J_1}^1
\end{equation}
annihilates $\psi^1_+$ making the superconnection a singlet under the full $su(1,1|3)$ algebra. As a consequence, defect operators (i.e. local operators of the theory inserted along the line) can be characterized by a set of four quantum numbers $(\D;m;j_1,j_2)$ associated to the 4 Cartan generators of the bosonic subalgebra (see appendix \ref{representations} for a classification of irreducible representations). 

On the fermionic side, 12 of the 24 original supersymmetry generators $Q^{IJ}_{\pm}$ and $S^{IJ}_{\pm}$ are preserved by the defect. Those are given by $\{Q_+^{1I},Q^{IJ}_-,S^{1I}_+,S^{IJ}_-\}$ for $I,J=2,3,4$, which we reorganize in (anti-)fundamental representations of $SU(3)$ as $\{Q^a,\bar Q_a, S^a, \bar S_a\}$ for $a=1,2,3$. R-symmetry and spinor indices have been raised and lowered with epsilon tensors as customary (see appendix \ref{conventions}).

Scalar and fermionic fields can be accommodated in the new R-symmetry pattern
\begin{align}
 C_I&=(Z, Y_a) &  \bar C^I&=(\bar Z, \bar Y^a)\\
 \psi_I^\pm&=(\psi^{\pm},\chi_a^\pm) &\bar \psi^I_\pm&=(\bar\psi_{\pm},\bar\chi^a_\pm)\label{fermions}
\end{align}
where $Y_a$ ($\bar Y^a$) and $\chi_a^\pm$ ($\bar\chi^a_\pm$) change in the $\mathbf{3}$ ($\mathbf{\bar 3}$) of $SU(3)$, whereas $Z$ and $\psi_+$ are singlet. Also the gauge fields can be split according to the new spacetime symmetry
\begin{equation}\label{gaugecomps}
 A_{\mu}=(A_1,A=A_2-i A_3,\bar A=A_2+i A_3)
\end{equation}
and similarly for $\hat{A}$.
In this notation the superconnection for the straight Wilson line simply reads
\begin{equation}
 \mathcal{L}=\mathcal{A}+\mathcal{L}_B+\mathcal{L}_F
\end{equation}
with
 \begin{align}\label{straightlineconn}
 \mathcal{A}&=\begin{pmatrix}  A_1  & 0\\
             0 & \hat A_1 
            \end{pmatrix} & \mathcal{L}_B&=\frac{2\pi i}{k}\begin{pmatrix}  Z \bar Z-Y_a \bar Y^a  & 0\\
             0 & \bar Z Z -\bar Y^a Y_a 
            \end{pmatrix} &  \mathcal{L}_F&=2\, \sqrt{\frac{\pi}{k}} \begin{pmatrix} 0  & - i\bar \psi_+ \\
            \psi^+ & 0
            \end{pmatrix}
\end{align}

\subsection{Defect correlation functions}
 As usual, symmetries put constraints on correlation functions. Here we focus on defect correlation functions, i.e. correlators of local operators inserted along the Wilson line. Since in this case the Wilson line is a $U(N|N)$ supermatrix, the natural insertion is a supermatrix $\mathcal{X}$, whose defect  two-point function is defined as
\begin{equation}\label{corrwithins}
 \braket{\mathcal{X}_1(\tau_1)\mathcal{X}_2(\tau_2)}_{\mathcal{W}}=\frac{\braket{\Tr P \mathcal{X}_1(\tau_1)\mathcal{W}(\t_1,\t_2)\mathcal{X}_2(\tau_2)\mathcal{W}(\t_2,\t_1)}}{\braket{\mathcal{W}}}
\end{equation}
with $\mathcal{W}(\t_1,\t_2)=P\exp\left(-i\int_{\tau_2}^{\tau_1} d\tau \mathcal{L}(\t) \right) $. Nevertheless, in the following we will use also two-point functions of objects changing in some representation of $U(N)\times U(N)$ instead of $U(N|N)$. These two-point functions has to be interpreted as the two-point function \eqref{corrwithins} with appropriate non-vanishing entries for the supermatrices $\mathcal{X}_1$ and $\mathcal{X}_2$. To give a specific example consider the two-point function $\braket{\psi^+(\tau_1) \bar \psi_+(\tau_2)}_{\mathcal{W}}$. The color indices of the fermions immediately indicate the possible position of $\psi^+(\tau_1)$ and $\bar\psi_+(\tau_2)$ inside the supermatrices, such that
\begin{equation}
 \braket{\psi^+(\tau_1) \bar \psi_+(\tau_2)}_{\mathcal{W}}=\braket{\begin{pmatrix}  0 & 0\\
             \psi^+(\tau_1) & 0
            \end{pmatrix}\begin{pmatrix}  0 & \bar \psi_+(\tau_2)\\
             0 & 0 
            \end{pmatrix}}_{\mathcal{W}}
\end{equation}
and the r.h.s. of \eqref{corrwithins} is now well-defined.

In the following we will be interested in defect two-point functions of local operators with classical dimension one. Those are fermions and scalar bilinears. The latter organize in irreducible representations of $SU(3)$ once the operator $C_I\bar C^J$ is properly decomposed\footnote{Remember that combinations involving two $C$'s or two $\bar C$'s are forbidden by the color structure}. In particular we have
 \begin{equation}
  \mathbf{4}\otimes\bar{\mathbf{4}}=\mathbf{1}\oplus\mathbf{1}\oplus\mathbf{3}\oplus\bar{\mathbf{3}}\oplus \mathbf{8}
 \end{equation}
Therefore we consider the five operators
\begin{equation}\label{scalarop}
 O_Z=Z \bar Z \quad O_Y=Y_a \bar Y^a \quad O_a=Y_a \bar Z \quad \bar O^a =Z \bar Y^a \quad O_a^b=Y_a \bar Y^b-\frac13 \d_a^b Y_c\bar Y^c 
\end{equation}
Notice that these operators change in the bifundamental representation of the first factor of the gauge group $U(N)\times U(N)$, therefore we have also a mirror set of operators $\{\hat O_Z,\hat O_Y,\hat O_a,\hat O^a,\hat O_a^b\}$ where the order of the factors in \eqref{scalarop} is exchanged. The kinematical part of the two-point functions of these operators are fixed by the residual symmetry. We list them here for the straight line case
\begin{equation}\label{boscorr}
 \braket{O_a(s_1) \bar O^b(s_2)}_{\mathcal{W}}=c_s \frac{\d_a^b}{{s_{12}}^{2}} \qquad  \braket{O_a^b(s_1) O_c^d(s_2)}_{\mathcal{W}}=k_s\frac{\d^b_c \d_a^d-\frac13\d_a^b \d_c^d}{{s_{12}}^{2\D_O}}
\end{equation}
where $s_{12}=s_1-s_2$ and we neglected the correlators involving singlet operators $O_Y$ and $O_Z$ since in general they mix and they are not important in the following. Moreover we assumed that the operator $O_a$ has conformal dimension one at the quantum level. In section \ref{dispmultiplet} we will show that $O_a$ is the highest weight operator of a 1/3 BPS multiplet containing the displacement operator, which guarantees that its dimension is protected from quantum corrections.

Let us analyse also fermionic operators. Notice that, despite descending from the three-dimensional spinors, the fermions in the defect theory (equivalent in this respect to a one-dimensional CFT) do not carry any spinor index and their correlation functions are similar to the scalar case, but for their Grassmann nature\footnote{In the derivation of the correlation functions we use also that the theory is parity invariant, which prevents us from easily extending these arguments to ABJ theory}. Notice that fermions, unlike the scalar bilinears \eqref{scalarop}, are charged also under $U(1)_M$ which prevents a coupling between $+$ and $-$. Therefore we are left with
\begin{align}\label{fercorr1}
 \braket{\psi^+(s_1) \bar \psi_+(s_2)}_{\mathcal{W}}& =i\, k_f \frac{s_{12}}{|s_{12}|^{2\D_{\psi^+}+1}} & \braket{\psi^-(s_1) \bar \psi_-(s_2)}_{\mathcal{W}}&=i\, \tilde{k}_f\frac{s_{12}}{|s_{12}|^{2\D_{\psi^-}+1}}\\
 \braket{\chi_a^+(s_1) \bar \chi^b_+(s_2)}_{\mathcal{W}}& =i\, c_f \frac{s_{12} \d_a^b}{|s_{12}|^{3}} & \braket{\chi_a^-(s_1) \bar \chi^b_-(s_2)}_{\mathcal{W}}&=i\, \tilde{c}_f \frac{s_{12} \d_a^b}{|s_{12}|^{2\D_{\chi^-}+1}}\label{fercorr2}
\end{align}
where the factors of $i$ are purely conventional. In this case we didn't indicate the conformal dimension of $\chi^+_a$ implying that it is protected. In section \ref{dispmultiplet} we will show that $\chi^+_a$ is part of a 1/2 BPS multiplet containing the fermionic part of the displacement operator.

All the arguments in this section have been carried out explicitly for the straight Wilson line with parametrization \eqref{simplepar}. Similar result can be derived for the cylinder parametrization \eqref{cylcoord} and for the circular Wilson line by taking the appropriate conformal mapping. The former case is particularly convenient if the two points sit on the same branch of the cusp in figure \ref{fig:cusp}. In this case the corresponding correlation functions can be simply obtained by the formal replacement
\begin{equation}\label{repcyl}
 s_{12}\to \sqrt{2(\cosh \tau_{12}-1)}
\end{equation}
For the circular case, on the other hand, one also needs to replace the fermions with appropriate eigenstates of $\dot{x}^{\mu} \g_{\mu}$ (for the line $\dot{x}^{\mu} \g_{\mu}$ is simply $\g_1$ and the label $\pm$ on fermions refers exactly to eigenstates of $\g_1$). Those are given by 
\begin{align}\label{upanddown1}
 \psi^{\uparrow}&=\frac{1}{\sqrt{2}}(e^{-i\frac{\tau}{2}} \psi^1+ i e^{i\frac{\tau}{2}} \psi^2) & \psi^{\downarrow}&=\frac{1}{\sqrt{2}}(e^{-i\frac{\tau}{2}} \psi^1+ i e^{i\frac{\tau}{2}} \psi^2) \\
 \bar \psi_{\uparrow}&=\frac{1}{\sqrt{2}}(e^{i\frac{\tau}{2}} \bar\psi_1- i e^{-i\frac{\tau}{2}} \bar \psi_2) & \bar \psi_{\downarrow}&=\frac{1}{\sqrt{2}}(e^{i\frac{\tau}{2}} \bar\psi_1+ i e^{-i\frac{\tau}{2}} \bar \psi_2) \label{upanddown2}
\end{align}
and similarly for $\chi$. Given these identifications one can find the correlation functions on the circle starting by \eqref{boscorr}, \eqref{fercorr1} and \eqref{fercorr2} and performing the formal replacements
\begin{equation}\label{repcirc}
 s_{12}\to \sqrt{2(1-\cos \tau_{12})} \qquad +\to\,  \uparrow \quad -\to \, \downarrow
\end{equation}

\section{Bremsstrahlung function and circular Wilson loop}
In the following the main  goal  is to prove a connection between the Bremsstrahlung function for the 1/2 BPS Wilson cusp and the circular Wilson loop, whose exact result is accessible, at least in principle, to supersymmetric localization.
The derivation goes along the lines of the four-dimensional case \cite{Correa:2012at}, with the notable complication of the fermionic degrees of freedom in the superconnection. We start by deriving an expression for the Bremsstrahlung function in terms of two-point functions of operators inserted along the line.

\subsection{Bremsstrahlung function and two-point functions}\label{Band2pt}
We consider the generalized cusp configuration described in section \ref{WLABJM} and we set to zero the physical angle $\varphi$. Thanks to the condition \eqref{smallangle}, the second-order expansion for small $\theta$ gives the Bremsstrahlung function. Therefore we consider the double derivative of the Wilson line expectation value \eqref{WL} with couplings \eqref{boslinecoupling} and \eqref{ferlinecoupling}. This simply gives
\begin{equation}
\frac12\left. \frac{\pa^2}{\pa \theta^2} \log\braket{\mathcal{W_\theta}}\right|_{\theta=0}=
 -\frac{1}{2N} \int_{-\infty}^{\infty} d\tau_1\int_{-\infty}^{\tau_1} d\tau_2 \braket{\mathcal{L}^{(1)}(\tau_1)\mathcal{L}^{(1)}(\tau_2)}_{\mathcal{W}_0}
\end{equation}
where we used the cylinder parametrization \eqref{cylcoord} and both points $\tau_1$ and $\tau_2$ are on one of the two antiparallel lines in figure \ref{fig:cusp} (or equivalently on one branch of the cusp). We indicated with $\mathcal{L}^{(1)}(\tau)$ the first-order expansion of the superconnection for small $\theta$
\begin{equation}
 \mathcal{L}(\tau)=\mathcal{L}^{(0)}(\tau)+\theta \mathcal{L}^{(1)}(\tau)+\mathcal{O}(\theta^2)
\end{equation}
Notice that we stopped the expansion at the first order since the second order would be related to one-point functions of local operators which vanish for the residual conformal invariance.
The explicit expression of $\mathcal{L}^{(1)}(\tau)$ is
\begin{equation}\label{L1line}
 \mathcal{L}^{(1)}=\begin{pmatrix} 
			    - \frac{2\pi i}{k} \left(O_1+\bar O^1\right) & i \sqrt{\frac{\pi}{k}} \bar \chi^1_+\\
			    - \sqrt{\frac{\pi}{k}} \chi_1^+  & -\frac{2\pi i}{k} \left( \hat O_1+ \hat{\bar O}^1\right) 
                         \end{pmatrix}
\end{equation}
with the operators defined in \eqref{scalarop} and \eqref{fermions}. Taking the products and using the properties of the correlation functions \eqref{boscorr}, \eqref{fercorr1} and \eqref{fercorr2} we find
\begin{align} 
-\braket{(\mathcal{L}^{(1)}(\tau_1)\mathcal{L}^{(1)}(\tau_2))}_{\mathcal{W}_0}&=\frac{8\pi^2}{k^2}\left(\braket{  O_1(\tau_1)\bar O^1(\tau_2) }_{\mathcal{W}_0}+\braket{ \hat O_1(\tau_1)\hat {\bar O}^1(\tau_2)}_{\mathcal{W}_0}\right)\nonumber \\
&+\frac{2\pi i}{k}\braket{ \bar \chi^1_+(\tau_1)\chi_1^+(\tau_2)}_{\mathcal{W}_0} \label{twopointOandchi}
\end{align}
Now we simply need to use \eqref{boscorr} and \eqref{fercorr2} after the replacement \eqref{repcyl} and keep in mind that $\tau_1>\tau_2$. This gives
\begin{equation}\label{finintline}
 \frac12\left. \frac{\pa^2}{\pa \theta^2} \log\braket{\mathcal{W_\theta}}\right|_{\theta=0}=\frac{1}{N}\left(\frac{4\pi^2}{k^2}(c_s+\hat c_s)-\frac{\pi}{k} c_f\right) \int_{-\infty}^{\infty} d\tau_1\int_{-\infty}^{\tau_1} d\tau_2 \frac{1 }{2(\cosh\tau_{12}-1)}
\end{equation}
The resulting integral is identical to the four-dimensional case \cite{Correa:2012at} and we can follow the same steps. We symmetrize the contour, factor out an overall divergence $T=\int_{-\infty}^{\infty}d\tau$ and perform the last integral
\begin{align}
  \left(\int_{-\infty}^{-\e} +\int_{\e}^{\infty}\right) d\tau \frac{1}{2(\cosh \t-1)}=-1
\end{align}
with a cut-off regularization and discarding power-law divergences. This leads to
\begin{equation}
 \frac12\left. \frac{\pa^2}{\pa \theta^2} \log\braket{\mathcal{W_\theta}}\right|_{\theta=0}=-\frac{T}{2N}\left(\frac{4\pi^2}{k^2}(c_s+\hat c_s)-\frac{\pi}{k} c_f\right)
\end{equation}
Comparing with \eqref{smallangle} and \eqref{Gammacusp} we conclude that
\begin{equation}\label{Bremtwopoint}
 B= \frac{1}{2N}\left(\frac{4\pi^2}{k^2}(c_s+\hat c_s) -\frac{\pi}{k} c_f\right) 
\end{equation}

\subsection{Two-point functions and circular Wilson line}
Since we derived an expression for the Bremsstrahlung function in terms of scalar and fermion two-point functions, we would like to relate such two-point functions to some circular Wilson loop. We therefore consider the latitude Wilson loop with couplings \eqref{latcoupling} and we take the derivative
\begin{equation}
 \left.\frac{\pa}{\pa \nu} \log\braket{\mathcal{W}_{\nu}}\right|_{\nu=1}=-\left. \frac{\pa^2}{\pa \theta_0^2} \log\braket{\mathcal{W_{\nu}}}\right|_{\theta_0=0}
\end{equation}
As for the linear case we start by expanding the superconnection at small $\theta_0$ (we set $\alpha=\frac{\pi}{4}$) 
\begin{equation}
 \mathcal{L}(\tau)=\mathcal{L}^{(0)}(\tau)+\theta_0 \mathcal{L}^{(1)}(\tau)+\mathcal{O}(\theta_0^2)
\end{equation}
with
\begin{equation}
 \mathcal{L}^{(1)}(\tau)=\begin{pmatrix} 
			    -\frac{2\pi i}{k} \left(e^{i\tau}O_1+e^{-i\tau}\bar O^1\right) & i\sqrt{\frac{\pi}{k}}e^{i\tau}  \bar\chi^1_{\uparrow}\\
			    -\sqrt{\frac{\pi}{k}}e^{-i\tau} \chi_1^{\uparrow}  & -\frac{2\pi i}{k} \left(e^{i\tau} \hat O_1+e^{-i\tau} \hat{\bar O}^1\right) 
                         \end{pmatrix}
\end{equation}
where we used the definitions \eqref{upanddown1} and \eqref{upanddown2}. The matrix $\mathcal{T}$ in \eqref{latcoupling} can be safely taken at the value $\theta_0=0$ and it transforms the supertrace in \eqref{WL} into a trace yielding
\begin{equation}
\left. \frac{\pa^2}{\pa \theta_0^2} \log\braket{\mathcal{W_\nu}}\right|_{\theta_0=0}=-\frac1{N}\int_0^{2\pi} d\tau_1\int_0^{\tau_1} d\tau_2 \braket{\mathcal{L}^{(1)}(\tau_1)\mathcal{L}^{(1)}(\tau_2)}_{\mathcal{W}_1}
\end{equation}
Taking products and using the properties of the correlation functions \eqref{boscorr}, \eqref{fercorr1}, \eqref{fercorr2} after the replacements \eqref{repcirc}, \eqref{upanddown1} and \eqref{upanddown2} we get
\begin{align} 
-\braket{(\mathcal{L}^{(1)}(\tau_1)\mathcal{L}^{(1)}(\tau_2))}_{\mathcal{W}_1}&=\frac{8\pi^2}{k^2}\cos \t_{12}\left( \braket{  O_1(\tau_1)\bar O^1(\tau_2) }_{\mathcal{W}_1} + \braket{ \hat O_1(\tau_1)\hat{\bar O}^1 (\tau_2)}_{\mathcal{W}_1}\right)\nonumber \\
&+\frac{2\pi i}{k}\cos \t_{12}\braket{ \bar \chi^1_{\uparrow}(\tau_1)\chi_1^{\uparrow}(\tau_2)}_{\mathcal{W}_1} \label{listcont}
\end{align}
and, keeping in mind that $\tau_1>\tau_2$
\begin{equation}
 \left. \frac{\pa^2}{\pa \theta_0^2} \log\braket{\mathcal{W_\nu}}\right|_{\theta_0=0}=\frac{2}{N}\left(\frac{4\pi^2}{k^2}(c_s+\hat c_s)-\frac{\pi}{k} c_f\right)\int_{0}^{2\pi} d\tau_1\int_{0}^{\tau_1} d\tau_2\,  \frac{\cos \tau_{12} }{2(1-\cos\tau_{12})}
\end{equation}
As for the line case we find an integral that already appeared in the four-dimensional case \cite{Correa:2012at}. Again, we symmetrize the contour, factor out a factor $2\pi$ and solve the last integral
\begin{equation}
 \int_{\e}^{2\pi-\e} \frac{\cos \tau }{1-\cos\tau} =- 2\pi
\end{equation}
disregarding power-law divergences. The final result reads
\begin{equation}
 \left.\frac{\pa}{\pa \nu} \log\braket{\mathcal{W}_{\nu}}\right|_{\nu=1}=\frac{2\pi^2}{N}\left(\frac{4\pi^2}{k^2}(c_s+\hat c_s)-\frac{\pi}{k} c_f\right)
\end{equation}
Comparing with \eqref{Bremtwopoint} we conclude that
\begin{equation}
 B(\l,N)=\frac{1}{4\pi^2} \left.\frac{\pa}{\pa \nu} \log\braket{\mathcal{W}_{\nu}}\right|_{\nu=1}
\end{equation}
which is the main result of our analysis.

\section{The superdisplacement operator}
The excitation of a conformal field theory by the insertion of an extended probe (a defect) clearly breaks translation invariance. In particular, the stress tensor is no longer conserved and the usual conservation law needs to be modified by some additional terms localized on the defect. For the case at hand, the conformal defect is a Wilson line in three-dimensional space and, for particularly symmetric configurations, such as the straight line or the circle, coordinates can be split in parallel and orthogonal ones, as we did in \eqref{coordsplit}.
In these coordinates the stress tensor conservation law can be written as
\begin{equation}\label{Ward}
 \partial^{\mu} T_{\mu m}(x)=\d^2(x_{\perp}) \mathbb{D}_m(x^1)
\end{equation}
However, one has to put particular care in interpreting this equation. Indeed, as we will see, in our case the r.h.s. is a $U(N|N)$ supermatrix while the l.h.s. is a bosonic bulk operator. The puzzle is solved by remembering that equation \eqref{Ward} is meaningful only when both sides are inserted inside a correlation function. In that case, being the r.h.s. localized on the Wilson line by the delta function, its supermatrix structure is very natural and equation \eqref{Ward} is well defined. 

The operator on the r.h.s. of \eqref{Ward} is called \textit{displacement operator} and, thanks to this Ward identity, it accounts for the variation of an arbitrary correlation function when the shape of the defect undergoes a small deformation. More specifically, let us consider the deformation of a linear defect parametrized by $x^1(s)=s$ by a profile $\delta x^m(s)$. An immediate consequence of the Ward identity \eqref{Ward} is that a correlation function of arbitrary operators $\braket{X}_{\mathcal{W}+\d \mathcal{W}}$ taken in presence of the deformed Wilson line, at first order in the deformation reads
\begin{equation}
 \braket{X}_{\mathcal{W}+\d \mathcal{W}} = \int ds \braket{X \mathbb{D}_m(s)}_\mathcal{W} \d x^m(s) +\mathcal{O}(\d x^2)
\end{equation}
This expression can be extended to the limit when no additional field $X$ is present, i.e. for the expectation value of the Wilson line. In that case however the first order variation vanishes, as it involves a defect one-point function, and the first non-trivial contribution shows up at second order in the deformation
\begin{equation}\label{deform}
 \d\log\braket{\mathcal{W}}=\int_{s_1>s_2} ds_1 ds_2 \braket{\mathbb{D}_m(s_1) \mathbb{D}_n(s_2)}_\mathcal{W} \d x^m(s_1) \d x^n(s_2) +\mathcal{O}(\d x^3)
\end{equation}
For the case of interest here, we can immediately make an interesting observation. The form of the displacement operator for a Wilson line can be computed explicitly by exploiting the formula for the variation of a Wilson line
\begin{equation}
\frac{\d \log \braket{\mathcal{W}(s_1,s_2)}}{\d x^m(s)} =-i\frac{\braket{\mathcal{W}(s_1,s)\frac{\d \mathcal{L}(x)}{\d x^m(s)} \mathcal{W}(s, s_2)}}{\braket{\mathcal{W}(s_1,s_2)}}=-i \braket{\frac{\d \mathcal{L}(x)}{\d x^m(s)}}_\mathcal{W}
\end{equation}
from which one can immediately identify
\begin{equation}\label{DvarL}
 \mathbb{D}_m(s)=-i\frac{\d \mathcal{L}(x)}{\d x^m(s)}
\end{equation}
In the present case the connection is a supermatrix and consequently, as we anticipated, the displacement operator is also a supermatrix. For the straight line case we get
\begin{equation}
 \mathbb{D}_m=\mathcal{F}_{m1}+D_m(\mathcal{L}_B+\mathcal{L}_F)
\end{equation}
with the definitions \eqref{straightlineconn} and the covariant derivative $D_m$ taken with respect to the gauge part of the superconnection
\begin{align}
D_m X&=\pa_m X+i[\mathcal{A}_m,X ]  &  \mathcal{A}_m=\begin{pmatrix} A_m  & 0\\
             0 & \hat A_m
            \end{pmatrix}
\end{align}
The field strength supermatrix is given by
\begin{align}            
\mathcal{F}_{m1}&= \pa_m \mathcal{A}_1-\pa_1 \mathcal{A}_m+i[\mathcal{A}_m,\mathcal{A}_1]
\end{align}

The Ward identity \eqref{Ward}, relating the divergence of the bulk stress tensor with the displacement operator protects the conformal dimension of the latter from quantum corrections. Furthermore, the fact that $\mathbb{D}_m$ is a supermatrix does not affect the general arguments according to which its two-point function is fully fixed by the residual symmetry leading to 
\begin{equation}\label{disptwopt}
 \braket{\mathbb{D}_m(s_1) \mathbb{D}_n(s_2)}_{\mathcal{W}}=\frac{\d_{mn}\,  C_D}{|s_{12}|^4}
\end{equation}
As shown by the authors of \cite{Correa:2012at}, the coefficient $C_D$, non-trivial function of the parameters of the theory, for the case of the Wilson line is just the Bremsstrahlung function. More precisely they found that
\begin{equation}
 C_D=12\, B
\end{equation}
Their argument, which we shortly review, goes along the lines of our previous derivation of formula \eqref{Band2pt}. They implement a deformation of the straight line into a cusp by considering an infinitesimal variation $\d x^m(s)=\varphi\, s \d^m_2$ for $s>0$ and small $\varphi$. Inserting this into \eqref{deform} and using \eqref{disptwopt} one gets
\begin{align}\label{cuspdef}
\d_{\text{cusp}}\log \braket{W}=\varphi^2 C_D\int_0^{L} ds_1 \int_{0}^{s_1-\e} ds_2 \frac{s_2s_1}{s_{12}^4}
\end{align}
After a change of variable $s_i=e^{\t_i}$ with $\t_i\in \mathbb{R}$, which is perfectly equivalent to map the problem on the cylinder as in figure \ref{fig:cusp} (see also the end of appendix \ref{conventions}) they obtain
\begin{equation}
 \d_{\text{cusp}}\log \braket{W}=\varphi^2 C_D \int_{\t_1>\t_2} d\t_1 d\t_2 \frac{1}{4(\cosh\t_{12}-1)}
\end{equation}
which is again the integral \eqref{finintline}. Following the same steps below \eqref{finintline} one gets
\begin{equation}
 \Gamma_{\text{cusp}}=\varphi^2 C_D \frac12 \int_{-\infty}^{\infty} d\t \frac{1}{4(\cosh\t-1)}=-\frac{C_D}{12} \varphi^2
\end{equation}
which proves the relation between $C_D$ and $B$ for an arbitrary conformal field theory.

\subsection{The displacement supermultiplet}\label{dispmultiplet}
Given the surprising relation between the two-point function of a complicated operator like the displacement and that of simple operators like $O_1$ and $\chi_1^+$ it is natural to ask whether supersymmetry relates those operators in some way. This would also guarantee that the operators $O_1$ and $\chi_1^+$ are protected from quantum corrections, a fact that we tacitly assumed in our derivation. Therefore in this section we want to understand which $su(1,1|3)$ supermultiplet the displacement operator belongs to. In appendix \ref{representations} we spell out short and long representations of $su(1,1|3)$, labelling them with the four Dynkin labels $\{\D, m, j_1, j_2\}$ of the highest weight state, as we pointed out in section \ref{symmetry}. The displacement operator has a free index in the orthogonal directions and it is convenient to separate it into two components with definite quantum numbers for rotations around the line. To do this we define
\begin{equation}
 \mathbb{D}=\mathbb{D}_2-i \mathbb{D}_3 \qquad \qquad  \bar{\mathbb{D}}=\mathbb{D}_2+i \mathbb{D}_3
\end{equation}
such that
\begin{equation}
 [i M_{23}, \mathbb{D}]=\mathbb{D} \qquad \qquad [i M_{23}, \bar {\mathbb{D}}]=-\bar{\mathbb{D}}
\end{equation}
Similarly we define
\begin{equation}\label{fieldstrenghtcomps}
 \mathcal{F}=\mathcal{F}_{21}-i \mathcal{F}_{31} \qquad  \bar{\mathcal{F}}=\mathcal{F}_{21}+i \mathcal{F}_{31}
\end{equation}
Notice that the previous operators are associated to the complex coordinates $z$ and $\bar z$.
After this recombination, we can assign to the two components of the displacement operator definite $su(1,1|3)$ quantum numbers
\begin{equation}
 \mathbb{D}\to \{2,3,0,0\} \qquad \qquad \bar{\mathbb{D}}\to \{2,-3,0,0\}
\end{equation}
By means of the equations of motion \eqref{Feqmot} we can eliminate the field strengths appearing in 
 the displacement operator in favour of scalar and fermion currents:
\begin{equation}\label{DBDF}
 \mathbb{D}=\mathbb{D}_B+\mathbb{D}_F
\end{equation}
with
\begin{align}\label{DBexpl}
 \mathbb{D}_B&=\small\frac{4\pi i}{k} \begin{pmatrix} ZD\bar Z- DY_a \bar Y^a+\bar\psi_+\psi^- +\bar\chi_+^a\chi^-_a & 0\\
            0& D\bar Z Z- \bar Y^a DY_a -\psi^-\bar\psi_+ -\chi^-_a\bar\chi_+^a
            \end{pmatrix} \\\normalsize \mathbb{D}_F&=2\, \sqrt{\frac{\pi}{k}} \begin{pmatrix} 0  & - iD\bar \psi_+ \\
            D\psi^+ & 0
            \end{pmatrix}\label{DFexpl}
\end{align}
and similarly for $\bar{\mathbb{D}}$.
Given these expressions we can locate any of the entries of the displacement supermultiplet in the appropriate $su(1,1|3)$ multiplet. Four good candidates are listed in appendix \ref{dispmult}. To select the appropriate ones, we could follow  \cite{Liendo:2016ymz} and impose the condition that every supersymmetry transformation on the displacement operator would yield a conformal descendant. However, the present case involves several complications. First of all, the preserved supersymmetry transformations, as a consequence of \eqref{susyvarL}, do not annihilate the super-holonomy, but only  its  supertrace. Secondly, the supermatrix nature of the displacement operator prevents from imposing strong conditions on the single entries. Despite these difficulties, let us attempt to derive a condition on the supersymmetry transformation of the displacement operator. We start from 
\begin{equation}
 \int d\tau \braket{\mathbb{D}(\tau) O_1(x_1)\dots O_n(x_n)}_{\mathcal{W}}=-\sum_{i=1}^n \pa_{x_i}\braket{O_1(x_1)\dots O_n(x_n)}_{\mathcal{W}} 
\end{equation}
and we assume, for notational simplicity, that all the operators $O_i$ are in the bulk. This has no influence on the final result. By taking the supersymmetry variation of the previous equation we find
\begin{equation}\label{Qondisp}
 \int d\tau \braket{\d_{\text{susy}}\left(\Tr[\mathcal{W}(+\infty,\tau) \mathbb{D}(\tau)\mathcal{W}(\tau,-\infty)]\right) O_1(x_1)\dots O_n(x_n)}=0
\end{equation}
where we made explicit the operator insertion \eqref{corrwithins}. Using
\begin{equation}\label{dandD}
 \pa_\tau \braket{\mathcal{O}(\tau)\dots}_{\mathcal{W}}= \braket{\mathcal{D}_\tau\mathcal{O}(\tau)\dots}_{\mathcal{W}}
\end{equation}
with the covariant derivative defined in \eqref{susyvarL}, we find that, for \eqref{Qondisp} to be true, we need to have
\begin{equation}\label{varD}
 \d_{\text{susy}}\left(\Tr[\mathcal{W}(+\infty,\tau) \mathbb{D}(\tau)\mathcal{W}(\tau,-\infty)]\right)=\Tr[\mathcal{W}(+\infty,\tau) \mathcal{D}_\tau\mathcal{O}(\tau)\mathcal{W}(\tau,-\infty)]
\end{equation}
This expression can be further simplified by noticing that \eqref{susyvarL} implies
\begin{equation}
 \d_{\text{susy}}\mathcal{W}(\tau_1,\tau_2)=i[\mathcal{W}(\tau_1,\tau_2)\mathcal{G}(\tau_2)-\mathcal{G}(\tau_1)\mathcal{W}(\tau_1,\tau_2)]
\end{equation}
Using this variation in \eqref{varD} and imposing that $\mathcal{G}$ vanishes at infinity (which is equivalent to ask that the straight infinite Wilson line is invariant under the preserved supersymmetries) we obtain 
\begin{equation}\label{susyvarD}
\d_{\text{susy}} \mathbb{D}(\tau)=\mathcal{D}_\tau \mathcal{O}(\tau)-i[\mathcal{G}(\tau),\mathbb{D}(\tau)]
\end{equation}
To find the explicit form of the operator $\mathcal{O}$ we use the explicit supersymmetry transformations in appendix \ref{susyonfields}. First we find that the supermatrix $\mathcal{G}$ in \eqref{susyvarL} reads
\begin{equation}\begin{split}
\mathcal{G}=4 \sqrt{\frac\pi k} \begin{pmatrix} 0  & \bar{\theta}^{a} Y_a \\
          -i\theta_a \bar{Y}^a & 0
            \end{pmatrix}
\end{split}\end{equation}
where the action of $Q^a$ and $\bar Q_a$ can be extracted by the differential operators $Q^a=\tfrac{\partial}{\partial\theta_a}$ and $\bar Q^a=\tfrac{\partial}{\partial\bar\theta^a}$. Then we can also see that
\begin{equation}
 \mathcal{O}(\tau)=-4 \sqrt{\frac\pi k} \begin{pmatrix} 2\sqrt{\frac{\pi}{k}} \bar{\theta}^a (Y_a \psi^- - Z \chi^-_a+\e_{abc} \bar\chi_+^b \bar{Y}^c)\quad  & -\bar{\theta}^{a} DY_a \\
          i\theta_a D\bar{Y}^a & 2\sqrt{\frac{\pi}{k}} \bar{\theta}^a ( \psi^- Y_a -  \chi^-_a Z+\e_{abc}  \bar{Y}^c \bar\chi_+^b)
            \end{pmatrix}
\end{equation}
The conjugate matrix would appear in the supersymmetry variation of $\bar{\mathbb{D}}$ 

The condition \eqref{susyvarD} is too weak to put constraints on the allowed $su(1,1|3)$ supermultiplets. Furthermore, constraining the whole supermatrix $\mathbb{D}$ and not the single components, its precise analysis would require combining the $su(1,1|3)$ multiplets into $U(N|N)$ representations, task which goes beyond the scope of this paper. Here we are only interested in identifying to which $su(1,1|3)$ multiplets the entries of \eqref{DBexpl} and \eqref{DFexpl} belong to. To do that we perform the explicit supersymmetry variation of the highest weights associated to the four candidate multiplets in appendix \ref{dispmult} and we locate the operators appearing in \eqref{DBexpl} and \eqref{DFexpl}. The details are given in appendix \ref{dispmult}. Here we only report the final result, reading
\begin{align}
 \mathbb{D}_F&\in \begin{pmatrix} 0  & \bar{\mathcal{B}}^{\frac12}_{\frac32,0,0} \\
            \bar{\mathcal{B}}^{\frac16}_{\frac52,0,1} & 0
            \end{pmatrix}  &
 \mathbb{D}_B&\in \begin{pmatrix} \bar{\mathcal{B}}^{\frac13}_{2,1,0}\oplus\bar{\mathcal{B}}^{\frac16}_{\frac52,0,1} & 0\\
            0& \bar{\mathcal{B}}^{\frac13}_{2,1,0}\oplus\bar{\mathcal{B}}^{\frac16}_{\frac52,0,1}
            \end{pmatrix}  
\end{align}
and conjugate ones for the $\bar{\mathbb{D}}$. Among the multiplets appearing in this result we notice the presence of $\bar{\mathcal{B}}^{\frac12}_{\frac32,0,0}$ and $\bar{\mathcal{B}}^{\frac13}_{2,1,0}$ which also include the operators  $\chi^+_a$ and $O^a=Z\bar Y^a$ respectively. Those are the same operators appearing in \eqref{twopointOandchi}, giving a further confirmation that they are protected from quantum corrections. Furthermore, the fact that they belong to the displacement supermultiplet may lead to a supersymmetry-based explanation for the fact that their defect two-point functions are closely related (more precisely the two-point function of $\mathcal{L}^{(1)}$ equals -- up to an overall coefficient-- that of $\mathbb{D}$, since they are both proportional to $B$). A precise understanding of this fact would require a precise analysis of the supersymmetric Ward identities, which we leave for future investigations.

As a last check of our derivation we now perform a perturbative computation of the Bremsstrahlung function, using \eqref{Bremtwopoint}.

\section{Perturbative checks}
Since both $c_f$ and $c_s$ in \eqref{Bremtwopoint} starts at order $k^0$ in a large $k$ expansion, the one-loop contribution to $B$ receives contributions only from the former ($c_s$ multiplies $\frac{1}{k^2}$ in \eqref{Bremtwopoint}), i.e. 
\begin{equation}
 B^{(1)}=-\frac{\pi}{2 k N}  c^{(0)}_f 
\end{equation}
The leading order for $c_f$ is easily extracted by comparing the correlators \eqref{fercorr2} with the propagator \eqref{ferpropdef} traced over color indices 
\begin{equation}
 c_f^{(0)}=-\frac{N^2}{4\pi}
\end{equation}
and this gives
\begin{equation}
 B^{(1)}=\frac{N}{8 k}
\end{equation}
in agreement with the literature.

At two loops the scalars start playing a role and we have 
\begin{equation}
 B^{(2)}= \frac{1}{2N}\left(\frac{4\pi^2}{k^2}(c^{(0)}_s+\hat c^{(0)}_s) -\frac{\pi}{k} c^{(1)}_f\right) 
\end{equation}
By looking at the scalar propagator \eqref{scalpropdef} we can easily extract 
\begin{equation}
c^{(0)}_s=\hat c^{(0)}_s=\frac{N^3}{16\pi^2} 
\end{equation}
The computation of $c_f^{(1)}$ is slightly more involved and here we anticipate the result of section \ref{twoloop}
\begin{equation}
c^{(1)}_f=\frac{N^3}{2\pi k} 
\end{equation}
yielding 
\begin{equation}
 B^{(2)}=0
\end{equation}
in agreement with the result of \cite{Griguolo:2012iq,Bianchi:2014laa,Bianchi:2017svd}.

\subsection{Computation of $c^{(1)}_f$}\label{twoloop}
We now give some details of the perturbative computation of $c^{(1)}_f$. We consider the fermionic part of $\mathcal{L}^{(1)}$ in \eqref{L1line}
\begin{equation}
 \L_F=\begin{pmatrix} 
			    0 & i \sqrt{\frac{\pi}{k}} \bar \chi_+^1\\
			    - \sqrt{\frac{\pi}{k}} \chi^+_1  & 0 
                         \end{pmatrix}
\end{equation}
and we compute its two-point correlation function on the straight line. This yields the value of $c^{(1)}_f$ since, for $s_1>s_2$ 
\begin{equation}
 \braket{\L_F(s_1)\L_F(s_2)}_{\mathcal{W}}=\frac{2\pi}{k}\frac{c_f}{{s_{12}}^2}
\end{equation}
Two classes of Feynman diagrams contribute to this two-point function\footnote{Since we are considering a fermion $\chi^+_a$ we don't have any coupling between the line and the fermionic part of the connection $\mathcal{L}_F$ which depends only on $\psi^+$.}\footnote{The diagrams and their related integrands are computed with the \textit{Mathematica}\textsuperscript{\textregistered} package \texttt{WiLE} \cite{Preti:2017fjb} with the algorithm slightly modified for the current computation.}. We consider them separately in the next two sections. We perform the computation employing dimensional regularization.

\subsubsection{Arcs}
\bigskip

\begin{center}
\begin{tikzpicture}[scale=0.7]
  \draw[blue,double,thick]  (-5,0)--(5,0);
  \draw[thick]  (3,0) arc (0:180:3);
  \draw[thick]  (1,0) arc (0:180:1);
  \node[below] at (-3,0) {$\L_F(s_2)$};
  \node[below] at (3,0) {$\L_F(s_1)$};
  \node[below] at (-1,0) {$\mathcal{L}_F(s_4)$};
  \node[below] at (1,0) {$\mathcal{L}_F(s_3)$};
 \end{tikzpicture}
 \begin{tikzpicture}[scale=0.7]
  \draw[blue,double,thick]  (-5,0)--(5,0);
  \draw[thick]  (3,0) arc (0:180:3);
  \draw[thick]  (1,0) arc (0:180:1);
  \node[below] at (-3,0) {$\mathcal{L}_F(s_4)$};
  \node[below] at (3,0) {$\mathcal{L}_F(s_3)$};
  \node[below] at (-1,0) {$\L_F(s_2)$};
  \node[below] at (1,0) {$\L_F(s_1)$};
 \end{tikzpicture}
 
 \bigskip
 
 \begin{tikzpicture}[scale=0.7]
  \draw[blue,double,thick]  (-5,0)--(5,0);
  \draw[thick]  (4,0) arc (0:180:1.5);
  \draw[thick]  (-1,0) arc (0:180:1.5);
 \node[below] at (1,0) {$\mathcal{L}_F(s_4)$};
  \node[below] at (4,0) {$\mathcal{L}_F(s_3)$};
  \node[below] at (-4,0) {$\L_F(s_2)$};
  \node[below] at (-1,0) {$\L_F(s_1)$};
 \end{tikzpicture}
 \begin{tikzpicture}[scale=0.7]
  \draw[blue,double,thick]  (-5,0)--(5,0);
  \draw[thick]  (4,0) arc (0:180:1.5);
  \draw[thick]  (-1,0) arc (0:180:1.5);
 \node[below] at (1,0) {$\L_F(s_2)$};
  \node[below] at (4,0) {$\L_F(s_1)$};
  \node[below] at (-4,0) {$\mathcal{L}_F(s_4)$};
  \node[below] at (-1,0) {$\mathcal{L}_F(s_3)$};
 \end{tikzpicture}
\end{center}
\medskip
In the first class of diagrams we have four different contributions which can be expressed as follows
\begin{align}
\left. \braket{\L_F(s_1)\L_F(s_2)}^{(1)}_{\mathcal{W}}\right|_{\text{arcs}} =&- \int_{s_2}^{s_1} ds_3 \int_{s_2}^{s_3} ds_4 \braket{\L_F (s_1) \mathcal{L}_F(s_3) \mathcal{L}_F(s_4) \L_F (s_2)} \nonumber\\
&- \int_{s_1}^{\infty} ds_3 \int_{-\infty}^{s_2} ds_4 \braket{\mathcal{L}_F(s_3)\L_F (s_1)   \L_F (s_2)\mathcal{L}_F(s_4)}\nonumber\\
&- \int_{s_1}^{\infty} ds_3 \int_{s_1}^{s_3} ds_4 \braket{ \mathcal{L}_F(s_3) \mathcal{L}_F(s_4)\L_F (s_1) \L_F (s_2)}\nonumber\\
&- \int_{-\infty}^{s_2} ds_3 \int_{-\infty}^{s_3} ds_4 \braket{\L_F (s_1)\L_F (s_2) \mathcal{L}_F(s_3) \mathcal{L}_F(s_4) }
\end{align}
where the matrix $\mathcal{L}_F$ is the fermionic part of the superconnection defined in \eqref{straightlineconn}.
It is straightforward to evaluate these diagrams by Wick contraction and using the propagators \eqref{ferpropdefdimreg}.
After performing the integrals and summing the four contributions the result reads
\begin{equation}
 \left. \braket{\L_F(s)\L_F(0)}^{(1)}_{\mathcal{W}}\right|_{\text{arcs}}=\frac{N^3}{2\pi^{1-2\e}k^2}\frac{ \Gamma(\frac12-\e)\Gamma(\frac32-\e)}{\e} \frac{(2L)^{2\e}+2 s^{2\e}}{ s^{2-2\e}} 
\end{equation}
This result is UV and IR divergent (the parameter $L$ is a long distance cut-off). Whereas the former divergence will be cancelled by the second class of diagrams, the latter should not be there. Nevertheless we should remember that, using dimensional regularization the result of $\braket{\mathcal{W}}$ is IR and UV divergent itself and the defect two-point point function, according to \eqref{corrwithins}, must be normalized by this factor. Therefore, the actual contribution of the arcs is
\begin{equation}\label{arcs}
\left. \braket{\L_F(s_1)\L_F(s_2)}^{(1)}_{\mathcal{W}}\right|_{\text{arcs}}-\braket{\L_F(s_1)\L_F(s_2)}^{(0)}\braket{\mathcal{W}}^{(1)}=\frac{N^3}{ k^2 }\frac{ \Gamma(\frac12-\e)\Gamma(\frac32-\e)}{\e} \left(\pi s^2\right)^{2\e-1} 
\end{equation}
The residual UV divergence will be cancelled by the vertices.

\subsubsection{Vertices}
\bigskip
\begin{center}
 \begin{tikzpicture}[scale=0.5]
  \draw[blue,double,thick]  (-5,0)--(5,0);
  \draw[thick]  (2,0) arc (0:180:2);
  \draw[decoration={aspect=0.3, segment length=1.55mm, amplitude=1.7mm,coil},decorate] (-1,1.71) arc (60:180:2);;
  \node[below] at (-2,0) {$\L_F(s_2)$};
  \node[below] at (2,0) {$\L_F(s_1)$};
  \node[below] at (-4,0) {$\mathcal{A}(s)$};
 \end{tikzpicture}
 \begin{tikzpicture}[scale=0.4]
  \draw[blue,double,thick]  (-5,0)--(5,0);
  \draw[thick]  (3,0) arc (0:180:3);
  \draw[decoration={aspect=0.3, segment length=1.55mm, amplitude=2mm,coil},decorate] (0,3)--(0,0);
  \node[below] at (-3,0) {$\L_F(s_2)$};
  \node[below] at (3,0) {$\L_F(s_1)$};
  \node[below] at (0,0) {$\mathcal{A}(s)$};
 \end{tikzpicture}
 \begin{tikzpicture}[scale=0.5]
  \draw[blue,double,thick]  (-5,0)--(5,0);
  \draw[thick]  (2,0) arc (0:180:2);
  \draw[decoration={aspect=0.3, segment length=1.55mm, amplitude=1.7mm,coil},decorate] (4,0) arc (0:120:2);;
  \node[below] at (-2,0) {$\L_F(s_2)$};
  \node[below] at (2,0) {$\L_F(s_1)$};
  \node[below] at (4,0) {$\mathcal{A}(s)$};
 \end{tikzpicture}
\end{center}

 In the second class of diagrams we have three different contributions which can be expressed as follows
 \begin{align}
\left. \braket{\L_F(s_1)\L_F(s_2)}^{(1)}_{\mathcal{W}}\right|_{\text{vertices}} =&-i \int_{-\infty}^{s_2} ds_3 \int d^3z \braket{\Tr[\L_F (s_1) \L_F (s_2)\mathcal{A}(s)] V_{\psi\psi A}(z)}\nonumber\\
&-i \int_{s_2}^{s_1} ds_3 \int d^3z \braket{\Tr[\L_F (s_1)\mathcal{A}(s) \L_F (s_2)] V_{\psi\psi A}(z)}\nonumber\\
&-i \int_{s_1}^{\infty} ds_3 \int d^3z \braket{\Tr[\mathcal{A}(s)\L_F (s_1) \L_F (s_2)] V_{\psi\psi A}(z)}
\end{align}
where the vertex $V_{\psi\psi A}(z)$ reads
\begin{equation}
 V_{\psi\psi A}(z)=\Tr[\bar \psi^I \gamma^{\mu}\psi_I A_{\mu}-\bar \psi^I \gamma^{\mu}\hat A_{\mu}\psi_I ]
\end{equation}
and the matrix $\mathcal{A}$ is defined in \eqref{straightlineconn}.
The final result reads 
\begin{equation}\label{vertices}
 \left. \braket{\L_F(s)\L_F(0)}^{(1)}_{\mathcal{W}}\right|_{\text{vertices}}=-\frac{N^3}{ k^2 }\frac{ \Gamma(\frac12-\e)\Gamma(\frac32-2\e)\Gamma(\e)\cos \pi \e}{2^{2\e}} \left(\pi s^2\right)^{2\e-1} 
\end{equation}
\subsubsection{Result}
Putting together \eqref{arcs} and \eqref{vertices} we obtain
\begin{equation}
 \braket{\L_F(s)\L_F(0)}^{(1)}_{\mathcal{W}}=\frac{N^3}{k^2\, s^2}
\end{equation}
yielding
\begin{equation}
 c^{(1)}_f=\frac{N^3}{2\pi k} 
\end{equation}
as expected.

\subsection*{Acknowledgements}
It is a pleasure to thank Marco Bianchi, Valentina Forini, Madalena Lemos, Pedro Liendo, Andrea Mauri, Marco Meineri, Carlo Meneghelli and Silvia Penati for very useful discussions and a critical reading of the draft.
The work of LB is supported by Deutsche Forschungsgemeinschaft in Sonderforschungsbereich 676 ``Particles, Strings, and the Early Universe. This work has also been supported in part by Italian Ministero dell'  Istruzione, Universit\`a e Ricerca (MIUR), Della Riccia Foundation and Istituto Nazionale di Fisica Nucleare (INFN) through the ``Gauge And String Theory'' (GAST) research project.  

\appendix
\section{Conventions}\label{conventions}
Our conventions for the spinor contractions are as follows
\begin{equation}
 \eta \bar \eta\equiv \eta^{\a} \bar \eta_{\a} \qquad \eta^{\a}=\e^{\a\b}\eta_{\b} \qquad \e^{\a\b} \e_{\b \g}=\d^\a_{\g} \qquad \e^{12}=-\e_{12}=1
\end{equation}
We work in Euclidean space $(x_1,x_2,x_3)$ with $\gamma$ matrices 
\begin{equation}
 {(\gamma^{\mu})_{\a}}^{\b}=(\s^1,\s^2,-\s^3)
\end{equation}
satisfying the Clifford algebra $\{\g^{\mu},\g^{\nu}\}=2\d^{\mu\nu} \mathbb{1}$. Notice that ${(\gamma^{\mu})^{\a}}_{\b}=\e^{\a\g}\e_{\b\d}{(\gamma^{\mu})_{\g}}^{\d}$, i.e.
\begin{equation}
 {(\gamma^{\mu})^{\a}}_{\b}=(\s^1,-\s^2,-\s^3)
\end{equation}
Since for the straight line we use the $\pm$ basis it is useful to also write down the gamma matrices in this basis
\begin{equation}
 {(\gamma^{\mu}_{\pm \text{ basis}})_\pm}^{\pm}=\{\s_3,\s_2,-\s_1\}
\end{equation}

The propagators for fundamental fields in 3d are 
\begin{align}
 \braket{{(C_I)_i}^{\hat j} (x) {(\bar C^J)_{\hat k}}^l (y)}^{(0)}&=\d_{I}^J \d_i^l \d_{\hat k}^{\hat j} \frac{1}{4\pi} \frac{1}{|x-y|} \\
 \braket{{(\psi^{\a}_I)_{\hat i}}^{j} (x) {(\bar \psi^J_{\b})_{k}}^{\hat l} (y)}^{(0)}&=\d_{I}^J \d_{\hat i}^{\hat l} \d_{ k}^{ j} \frac{-i}{4\pi} \frac{(x-y)_{\mu} ({\g^{\mu})_\b}^{\a}}{|x-y|^3}\\
 \braket{A_{\mu}^A(x)A_{\nu}^B(y)}^{(0)}&=\d^{AB} \frac{2\pi i}{k} \frac{1}{4\pi} \e_{\mu\nu\rho} \frac{(x-y)^{\rho}}{|x-y|^3}\\
  \braket{\hat{A}_{\mu}^A(x)\hat{A}_{\nu}^B(y)}^{(0)}&=-\d^{AB} \frac{2\pi i}{k} \frac{1}{4\pi} \e_{\mu\nu\rho} \frac{(x-y)^{\rho}}{|x-y|^3}
\end{align}
and in dimensional regularization
 \begin{align}
 \braket{{(C_I)_i}^{\hat j} (x) {(\bar C^J)_{\hat k}}^l (y)}^{(0)}&=\d_{I}^J \d_i^l \d_{\hat k}^{\hat j} \frac{\Gamma(\frac12-\epsilon)}{4\pi^{\frac32-\epsilon}} \frac{1}{|x-y|^{1-2\epsilon}} \\
 \braket{{(\psi^{\a}_I)_{\hat i}}^{j} (x) {(\bar \psi^J_{\b})_{k}}^{\hat l} (y)}^{(0)}&=\d_{I}^J \d_{\hat i}^{\hat l} \d_{ k}^{ j} \frac{-i\Gamma(\frac32-\epsilon)}{2\pi^{\frac32-\epsilon}} \frac{(x-y)_{\mu} ({\g^{\mu})_\b}^{\a}}{|x-y|^{3-2\e}}\\
 \braket{A_{\mu}^A(x)A_{\nu}^B(y)}^{(0)}&=\d^{AB} \frac{2\pi i}{k} \frac{\Gamma(\frac32-\epsilon)}{2\pi^{\frac32-\epsilon}} \e_{\mu\nu\rho} \frac{(x-y)^{\rho}}{|x-y|^{3-2\e}}\\
  \braket{\hat{A}_{\mu}^A(x)\hat{A}_{\nu}^B(y)}^{(0)}&=-\d^{AB} \frac{2\pi i}{k} \frac{\Gamma(\frac32-\epsilon)}{2\pi^{\frac32-\epsilon}} \e_{\mu\nu\rho} \frac{(x-y)^{\rho}}{|x-y|^{3-2\e}}
\end{align}
where $A,B$ are adjoint $SU(4)$ indices.

We summarize also our conventions on the straight line and circular 1/2 BPS Wilson lines. 
The contours are
\begin{align}
 &\text{Line} & x^{\mu}(\tau)&=(s,0,0) \label{simplepar}\\
 &\text{Circle} & x^{\mu}(\tau)&=(\cos\tau,\sin\tau,0)
\end{align}
with superconnection 
\begin{align}
\mathcal{L}&=\begin{pmatrix} A_\mu \dot{x}^\mu-\frac{2\pi i}{\k}  {M_J}^I C_I \bar C^J & -i\sqrt{\frac{2\pi}{\k}}\eta \bar \psi\\
             \sqrt{\frac{2\pi}{\k}} \psi \bar \eta & \hat A_\mu \dot x^\mu-\frac{2\pi i}{\k}  {M_J}^I  \bar C^J C_I
            \end{pmatrix}
\end{align} 
and
\begin{align}
 &\text{Line} & {M_I}^J&=\begin{pmatrix}
                    -1 	&0	&0	&0\\
                    0	&1		&0	&0\\
                    0		&0				&1	&0\\
                    0		&0				&0	&1\\
                   \end{pmatrix} &
          \eta^\a&=\frac{1}{\sqrt2} \begin{pmatrix}
                    1 & 1
                    \end{pmatrix}&
      \bar\eta_\a&=\frac{1}{\sqrt2} \begin{pmatrix}
                    1  \\ 1
                    \end{pmatrix}         
                   \\
 &\text{Circle} & {M_I}^J&=\begin{pmatrix}
                    -1 	&0	&0	&0\\
                    0	&1		&0	&0\\
                    0		&0				&1	&0\\
                    0		&0				&0	&1\\
                   \end{pmatrix} &
          \eta^\a&=\frac{1}{\sqrt2} \begin{pmatrix}
                    e^{i\frac{\t}{2}} & -ie^{-i\frac{\tau}2}
                    \end{pmatrix}&
      \bar\eta_\a&=\frac{1}{\sqrt2} \begin{pmatrix}
                    e^{-i\frac{\t}{2}} \\ ie^{i\frac{\tau}2}
                    \end{pmatrix}
\end{align}
The propagators for scalar fields inserted on the loop are
\begin{align}
&\text{Line} &\braket{{(C_I)_i}^{\hat j} (\tau_1) {(\bar C^J)_{\hat k}}^l (\tau_2)}_{\mathcal{W}}^{(0)}&=\d_{I}^J \d_i^l \d_{\hat k}^{\hat j} \frac{1}{4\pi} \frac{1}{|s_{12}|}\label{scalpropdef} \\
&\text{Circle} &\braket{{(C_I)_i}^{\hat j} (\tau_1) {(\bar C^J)_{\hat k}}^l (\tau_2)}_{\mathcal{W}}^{(0)}&=\d_{I}^J \d_i^l \d_{\hat k}^{\hat j} \frac{1}{4\pi} \frac{1}{2|\sin\frac{\t_{12}}{2}|}
\end{align}
and, in dimensional regularization
\begin{align}
&\text{Line} &\braket{{(C_I)_i}^{\hat j} (\tau_1) {(\bar C^J)_{\hat k}}^l (\tau_2)}_{\mathcal{W}}^{(0)}&=\d_{I}^J \d_i^l \d_{\hat k}^{\hat j}\frac{\Gamma(\frac12-\epsilon)}{4\pi^{\frac32-\epsilon}}\frac{1}{|s_{12}|^{1-2\epsilon}}\label{scalpropdefdimreg} \\
&\text{Circle} &\braket{{(C_I)_i}^{\hat j} (\tau_1) {(\bar C^J)_{\hat k}}^l (\tau_2)}_{\mathcal{W}}^{(0)}&=\d_{I}^J \d_i^l \d_{\hat k}^{\hat j} \frac{\Gamma(\frac12-\epsilon)}{4\pi^{\frac32-\epsilon}}\frac{1}{|2\sin\frac{\tau_{12}}{2}|^{1-2\epsilon}}
\end{align}
For fermions it is useful to consider the components
\begin{align}
&\text{Line} &\psi_I^+&=\psi_I \bar\eta & \bar\psi^I_+&=\eta\bar \psi^I\\
&\text{Circle} &\psi_I^{\uparrow}&=\psi_I \bar\eta & \bar\psi^I_{\uparrow}&=\eta\bar \psi^I
\end{align}
where we keep the $SU(4)$ indices since at tree level the symmetry breaking has no effect. Their propagators read
\begin{align}
 &\text{Line} & \braket{{(\psi^{+}_I)_{\hat i}}^{j} (s_1) {(\bar \psi^J_{+})_{k}}^{\hat l} (s_2)}_{\mathcal{W}}^{(0)}&=\frac{\d_I^J\d_{\hat i}^{\hat l} \d_{ k}^{ j}}{4\pi s_{12}|s_{12}|}\label{ferpropdef} \\
 &\text{Circle} & \braket{{(\psi^{\uparrow}_I)_{\hat i}}^{j} (\t_1) {(\bar \psi^J_{\uparrow})_{k}}^{\hat l} (\t_2)}_{\mathcal{W}}^{(0)}&=\frac{\d_I^J\d_{\hat i}^{\hat l} \d_{ k}^{ j}}{16\pi\sin\frac{\tau_{12}}{2}|\sin\frac{\tau_{12}}{2}|}
\end{align}
and, in dimensional regularization
\begin{align}
 &\text{Line} & \braket{{(\psi^{+}_I)_{\hat i}}^{j} (s_1) {(\bar \psi^J_{+})_{k}}^{\hat l} (s_2)}_{\mathcal{W}}^{(0)}&=\d_I^J \d_{\hat i}^{\hat l} \d_{ k}^{ j}\frac{\Gamma(\frac32-\epsilon)}{2\pi^{\frac32-\epsilon}}\frac{s_{12}}{|s_{12}|^{3-2\epsilon}}\label{ferpropdefdimreg} \\
 &\text{Circle} & \braket{{(\psi^{\uparrow}_I)_{\hat i}}^{j} (\tau_1) {(\bar \psi^J_{\uparrow})_{k}}^{\hat l} (\tau_2)}_{\mathcal{W}}^{(0)}&=\d_I^J \d_{\hat i}^{\hat l} \d_{ k}^{ j}\frac{\Gamma(\frac32-\epsilon)}{\pi^{\frac32-\epsilon}}\frac{\sin\frac{\tau_{12}}{2}}{|2\sin\frac{\tau_{12}}{2}|^{3-2\epsilon}}
\end{align}
Finally for the gauge field, the components can be split as in \eqref{gaugecomps} and the non-vanishing propagators are
\begin{align}
 &\text{Line} & \braket{A^A(s_1) \bar A^B(s_2)}_{\mathcal{W}}^{(0)}&=-\braket{\bar A^A(s_1)  A^B(s_2)}_{\mathcal{W}}^{(0)}=-\d^{AB} \frac{2\pi}{k} \frac{1}{4\pi} \frac{1}{s_{12}|s_{12}|} \\
 & &\braket{\hat A^A(s_1)  \hat{\bar A}^B(s_2)}_{\mathcal{W}}^{(0)}&=-\braket{\hat {\bar{A}}^A(s_1)  \hat A^B(s_2)}_{\mathcal{W}}^{(0)}=\d^{AB} \frac{2\pi}{k} \frac{1}{4\pi} \frac{1}{s_{12}|s_{12}|}\\
 &\text{Circle} & \braket{A^A(\t_1) \bar A^B(\t_2)}_{\mathcal{W}}^{(0)}&=-\braket{\bar A^A(\t_1)  A^B(\t_2)}_{\mathcal{W}}^{(0)}=-\d^{AB} \frac{2\pi}{k} \frac{1}{16\pi} \frac{1}{\sin\frac{\tau_{12}}{2}|\sin\frac{\tau_{12}}{2}|} \\
 & &\braket{\hat A^A(\t_1)  \hat{\bar A}^B(\t_2)}_{\mathcal{W}}^{(0)}&=-\braket{\hat {\bar{A}}^A(\t_1)  \hat A^B(\t_2)}_{\mathcal{W}}^{(0)}=\d^{AB} \frac{2\pi}{k} \frac{1}{16\pi} \frac{1}{\sin\frac{\tau_{12}}{2}|\sin\frac{\tau_{12}}{2}|}
\end{align}
In dimensional regularization they read
\begin{align}
 &\text{Line} & \braket{A^A(s_1) \bar A^B(s_2)}_{\mathcal{W}}^{(0)}&=-\d^{AB} \frac{2\pi}{k} \frac{\Gamma(\frac32-\epsilon)}{2\pi^{\frac32-\epsilon}} \frac{s_{12}}{|s_{12}|^{3-2\epsilon}} \\
 & &\braket{\hat A^A(s_1)  \hat{\bar A}^B(s_2)}_{\mathcal{W}}^{(0)}&\d^{AB} \frac{2\pi}{k} \frac{\Gamma(\frac32-\epsilon)}{2\pi^{\frac32-\epsilon}} \frac{s_{12}}{|s_{12}|^{3-2\epsilon}}\\
 &\text{Circle} & \braket{A^A(\t_1) \bar A^B(\t_2)}_{\mathcal{W}}^{(0)}&-\d^{AB} \frac{2\pi}{k} \frac{\Gamma(\frac32-\epsilon)}{\pi^{\frac32-\epsilon}} \frac{\sin\frac{\tau_{12}}{2}}{|2\sin\frac{\tau_{12}}{2}|^{3-2\epsilon}} \\
 & &\braket{\hat A^A(\t_1)  \hat{\bar A}^B(\t_2)}_{\mathcal{W}}^{(0)}&=\d^{AB} \frac{\pi}{k}\frac{\Gamma(\frac32-\epsilon)}{2\pi^{\frac32-\epsilon}}\frac{\sin\frac{\tau_{12}}{2}}{|2\sin\frac{\tau_{12}}{2}|^{3-2\epsilon}}
\end{align}

For the straight line, in section \ref{Band2pt}, we find it useful to map the problem on the cylinder $\mathbb{R}\times S^2$ with metric 
\begin{equation}
 dx^{\mu} dx_{\mu}=dr^2 + r^2 d\O_2=e^{2 t} (dt^2+d\O_2)
\end{equation}
The profile of the Wilson line in these coordinates is 
\begin{equation}\label{cylcoord}
\left\{ \begin{array}{l}
       t=\t  \\
	\varphi=0 
	\end{array}\right. \qquad \cup \qquad \left\{ \begin{array}{l}
        t=\t  \\
	\varphi=\pi 
        \end{array}\right.
\end{equation}
where the relation between $r$ and $t$ is the usual exponential map
\begin{equation}
 r=e^t
\end{equation}
and the logarithmic divergence for $r=\e$ and $r=L$ in \eqref{Gammacusp} maps to a linear divergence for $t=\pm \infty$ in \eqref{potential}.

\section{$osp(6|4)$ algebra}\label{algebra}
We now list the commutation relation for the $osp(6|4)$ superalgebra. Let us start from the three-dimensional conformal algebra
\begin{align}
[P^{\mu},K^{\nu}]&=2\d^{\mu\nu} D+2 M^{\mu\nu} & [D,P^{\mu}]&= P^{\mu} & [D,K^{\mu}]&=-K^{\mu}\\
 [M^{\mu\nu}, M^{\rho\sigma}]&=\d^{\sigma[\mu} M^{\nu]\rho}+\d^{\rho[\nu} M^{\mu]\sigma} & [P^{\mu},M^{\nu\rho}]&=\delta^{\mu[\nu}P^{\rho]} & [K^{\mu},M^{\nu\rho}]&=\delta^{\mu[\nu}K^{\rho]} 
\end{align}
Then we have the $SU(4)$ generators
\begin{align}
 [{J_I}^J,{J_K}^L]=\d_I^L {J_K}^J-\d^J_{K} {J_I}^L
\end{align}

Fermionic generators $Q^{IJ}_\a$ and $S^{IJ}_\a$ respect the reality condition $\bar Q_{IJ\a}=\frac12 \e_{IJKL} Q^{KL}_{\a}$ and similarly for $S$. Anticommutation relations are
\begin{align}
 \{Q^{IJ}_\a,Q^{KL\b}\}&=2\e^{IJKL} {(\gamma^\mu)_{\a}}^{\b} P_\mu \qquad  \{S^{IJ}_\a,S^{KL\b}\}=2\e^{IJKL} {(\gamma^\mu)_{\a}}^{\b} K_\mu \\  \{Q^{IJ}_\a,S^{KL\b}\}&=\e^{IJKL} ({(\gamma^{\mu\nu})_{\a}}^{\b} M_{\mu\nu}+2\d_{\a}^{\b} D)+2\d_{\a}^{\b}\e^{KLMN}(\d_M^J {J_N}^I-\d_M^I {J_{N}}^J)
\end{align}
Finally, mixed commutators are
\begin{align}
 [D,Q^{IJ}_{\a}]&=\frac12 Q^{IJ}_{\a} & [D,S^{IJ}_{\a}]&=-\frac12 S^{IJ}_{\a} \\
 [M^{\mu\nu},Q^{IJ}_{\a}]&=-\frac12 {(\gamma^{\mu\nu})_{\a}}^{\b} Q^{IJ}_{\b} & [M^{\mu\nu},S^{IJ}_{\a}]&=-\frac12 {(\gamma^{\mu\nu})_{\a}}^{\b} S^{IJ}_{\b}  \\
 [K^{\mu},Q_{\a}^{IJ}]&={(\g^{\mu})_{\a}}^{\b} S^{IJ}_{\b} & [P^{\mu},S^{IJ}_{\a}]&={(\g^{\mu})_{\a}}^{\b} Q^{IJ}_{\b}\\
 [{J_I}^J,Q^{KL}_{\a}]&=\d_I^K Q^{JL}_\a+\d_{I}^L Q^{KJ}_\a-\frac12\d_I^J Q^{KL}_{\a} &  [{J_I}^J,S^{KL}_{\a}]&=\d_I^K S^{JL}_\a+\d_{I}^L S^{KJ}_\a-\frac12\d_I^J S^{KL}_{\a}
\end{align}

\section{The subalgebra $su(1,1|3)$}\label{subalgebra}
Inside the $osp(6|4)$ it is possible to identify the $su(2|3)$ (or, more precisely $su(1,1|3)$) subalgebra preserved by the 1/2 BPS Wilson line. The $su(1,1)$ generators are those of the one-dimensional conformal group, i.e. $\{D,P_1,K_1\}$. Since for building irreducible representations it will be important to choose the correct real section, compared to the previous section we make the transformations $P_1\to i P_1$ and $K_1\to i K_1$ in order to obtain the correct $su(1,1)$ commutation relations
\begin{align}
 [P_1,K_1]&=-2 D & [D,P_1]&=P_1 & [D,K_1]&=-K_1
\end{align}
The $su(3)$ generators ${R_a}^b$ are traceless, i.e. ${R_a}^a=0$ and they are given in terms of the original $su(4)$ ones by
\begin{align}
{R_{a}}^{b}&=\begin{pmatrix}
                             {J_2}^2+\frac13 {J_1}^1 & {J_2}^3 & {J_2}^4\\
                             {J_3}^2 & {J_3}^3+\frac13 {J_1}^1 & {J_3}^4\\
                              {J_4}^2 & {J_4}^3  & -{J_3}^3-{J_2}^2-\frac23 {J_1}^1\\
                            \end{pmatrix}             
\end{align}
Their commutation relations are
\begin{align}
 [{R_{a}}^{b},{R_{c}}^{d}]&=\d_{a}^{d} {R_{c}}^{b} -\d^{b}_{c} {R_{a}}^{d}
\end{align}
The last bosonic symmetry is the $u(1)$ algebra generated by
\begin{align}
M=3i M_{23}-2 {J_1}^1
\end{align}
and commuting with the other bosonic generators.

The fermionic generators are given by a reorganization of the preserved supercharges\newline $\{Q^{12}_+,Q^{13}_+,Q^{14}_+,Q^{23}_-,Q^{24}_-,Q^{34}_-\}$, together with the corresponding superconformal charges. Our notation is
\begin{equation}
 Q^a=Q^{1a}_{+} \qquad S^a=i\, S^{1a}_{+}  \qquad  \bar Q_a=i\,\frac12 \epsilon_{abc} Q_{-}^{bc} \qquad  \bar S_a=\frac12 \epsilon_{abc} S_{-}^{bc}
\end{equation}
The $i$ factors are chosen to compensate the transformations on $P_1$ and $K_1$ so that anticommutators read
\begin{align}
 \{Q^a,\bar Q_b\}&=2 \d^a_b P_1 &  \{S^a,\bar S_b\}&=2\d^a_b K_1 \\  \{Q^a,\bar S_b\}&= 2\d_b^a ( D+\tfrac13 M)-2 {R_b}^a & \{\bar Q_a, S^b\}&= 2\d_b^a(D-\tfrac13 M)+2 {R_a}^b \label{anticommQS}
\end{align}
Finally, non-vanishing mixed commutators are\small
\begin{align}
 [D,Q^a]&=\frac12 Q^a & [D,\bar Q_a]&=\frac12 \bar Q_a &  [K_1,Q^a]&=S^a & [K_1,\bar Q_a]&= \bar S_a\\
 [D,S^a]&=-\frac12 S^a & [D,\bar S_a]&=-\frac12 \bar S_a &  [P_1,S^a]&=-Q^a & [P_1,\bar S_a]&=- \bar Q_a\\
 [{R_a}^b,Q^c]&=\d_a^c Q^b-\tfrac13 \d_a^b Q^c & [{R_a}^b,\bar Q_c]&=-\d_c^b \bar Q_a+\tfrac13 \d_a^b \bar Q_c & [M,Q^a]&=\tfrac12 Q^a & [M,\bar Q^a]&=-\tfrac12 \bar Q^a\\
 [{R_a}^b,S^c]&=\d_a^c S^b-\tfrac13 \d_a^b S^c & [{R_a}^b,\bar S_c]&=-\d_c^b \bar S_a+\tfrac13 \d_a^b \bar S_c & [M,S^a]&=\tfrac12 S^a & [M,\bar S^a]&=-\tfrac12 \bar S^a
\end{align}\normalsize

\subsection{Representations of $su(1,1|3)$}\label{representations}
We consider here long and short multiplets of the $su(1,1|3)$ algebra. The algebra is characterized by four Dynkin labels $[\D,m,j_1,j_2]$ associated to the Cartan generators of the bosonic subalgebra $su(1,1)\oplus u(1)\oplus su(3)$. With respect to our previous conventions it is convenient to rewrite the $su(3)$ generators in a Cartan-Weyl basis. We consider the Cartan subalgebra generated by
\begin{equation}
 J_1={R_1}^1-{R_2}^2 \qquad J_2={R_1}^1+2{R_2}^2
\end{equation}
and we relabel $E^+_a={R_{a+1}}^a$ and $E^-_a={R_a}^{a+1}$ with the sum on the indices performed modulo 3. This way we have a $su(2)$ subalgebra associated to any Cartan generator
\begin{equation}
 [J_a,E^{\pm}_a]=\pm 2E^{\pm}_a \qquad [E^+_a,E^-_a]=J_a
\end{equation}
with the identification $J_3=J_1+J_2$. Furthermore we have 
\begin{align}
 [E^{\pm}_1,E^{\pm}_2]&=E^{\pm}_3 & [E^{\pm}_{1,2},E^{\pm}_3]&=0&
 [J_1,E^{\pm}_2]&=\mp E^{\pm}_2 & [J_2,E^{\pm}_1]&=\mp E^{\pm}_1
\end{align}
In this basis the supercharges have definite quantum numbers and their action on a state $\ket{\D,m,j_1,j_2}$ can be simply obtained by shifts in the labels. In particular the associated charges are
\begin{align}
 Q^1\quad &[\tfrac12, \tfrac12, 1,0] & Q^2\quad  &[\tfrac12, \tfrac12, -1,1] & Q^3 \quad  &[\tfrac12, \tfrac12, 0,-1]\\
 \bar Q_1\quad &[\tfrac12, -\tfrac12, -1,0] & \bar Q_2\quad  &[\tfrac12, -\tfrac12, 1,-1]  & \bar Q_3\quad  &[\tfrac12, -\tfrac12, 0,1]
\end{align}
We can also list the charges of the fundamental fields of the theory
\begin{align}\label{fieldcharges1}
 Z\quad &[\tfrac12, \tfrac32, 0,0] & \bar Z\quad  &[\tfrac12, -\tfrac32, 0,0] & \bar Y^1 \quad  &[\tfrac12, \tfrac12, 1,0] &  Y_3 \quad  &[\tfrac12, -\tfrac12, 0,1]\\
  \psi^{+}\quad &[1, 0, 0,0] & \bar  \psi_{+}\quad  &[1, 0, 0,0] & \chi^+_3\quad  &[1, -2, 0,1] & \bar \chi_+^1\quad  &[1, 2, 1,0]\label{fieldcharges2}\\
 \psi^{-}\quad &[1, 3, 0,0] & \bar \psi_-\quad  &[1, -3, 0,0] & \chi^-_3\quad  &[1, 1, 0,1] & \bar \chi_-^1\quad  &[1, -1, 1,0]\label{fieldcharges3}
\end{align}
where we listed only the R-symmetry highest weights since the rules for different indices are identical to the ones for supercharges. We now proceed with the construction of the multiplets.

The long multiplet can be easily built by acting with the supercharges $Q^a$ and $\bar Q_a$ and the operators $E^-_a$ and $P_1$ on a highest weight state characterized by
\begin{align}
 S^a\ket{\D,m,j_1,j_2}^{\text{hw}}&=0 & \bar S_a\ket{\D,m,j_1,j_2}^{\text{hw}}&=0 & E^+_a\ket{\D,m,j_1,j_2}^{\text{hw}}&=0
\end{align}
The dimension of this module is 
\begin{equation}
 \text{dim} \mathcal{A}^{\D}_{m;j_1,j_2}=27 (j_1+1)(j_2+1)(j_1+j_2+2)
\end{equation}
and unitarity requires
\begin{equation}
 \D\geq\left\{\begin{array}{l}
           \frac13(2j_1+j_2-m) \quad m\leq \frac{j_1-j_2}{2}\\
           \frac13(j_1+2j_2+m) \quad m > \frac{j_1-j_2}{2}
          \end{array}\right.
\end{equation}

We now consider the possible shortening conditions one can get. Let us start by the multiplets of the kind $\mathcal{B}_{m;j_1,j_2}$ obtained imposing
\begin{equation}
 Q^a\ket{\D,m,j_1,j_2}^{\text{hw}}=0
\end{equation}
for the three cases  
\begin{align}
 a&=1 & \D&=\frac13(2 j_1+ j_2-m) & &\mathcal{B}^{\frac16}_{m,j_1,j_2}\\
 a&=1,2 & \D&=\frac13(j_2-m)  \quad j_1=0 & &\mathcal{B}^{\frac13}_{m,0,j_2}\\
 a&=1,2,3 & \D&=-\frac13 m \qquad \quad \, j_1=j_2=0 & &\mathcal{B}^{\frac12}_{m,0,0}
\end{align}
where, as usual, the conditions on the dimensions have been imposed by looking at consistency with the anticommutation relation \eqref{anticommQS}. We labelled these three multiplets according to the fraction of supercharges annihilating the highest weight state. 

The conjugate pattern emerges for the case of $\bar Q_a$. We have
\begin{equation}
 \bar Q_a\ket{\D,m,j_1,j_2}^{\text{hw}}=0
\end{equation}
for the three cases  
\begin{align}
 a&=3 & \D&=\frac13 ( j_1+2j_2+m) & &\bar{\mathcal{B}}^{\frac16}_{m,j_1,j_2}\\
 a&=2,3 & \D&=\frac13 ( j_1+m) \quad j_2=0 & &\bar{\mathcal{B}}^{\frac13}_{m,j_1,0}\\
 a&=1,2,3 & \D&=\frac13m \qquad \qquad  j_1=j_2=0 & &\bar{\mathcal{B}}^{\frac12}_{m,0,0}
\end{align}

Finally we may have mixed multiplets where the highest weight is annihilated both by $Q^a$ and $\bar Q_a$. Those include
\begin{align}
 &\hat{\mathcal{B}}^{\frac16 \frac16}_{m,j_1,j_2} & \D&=\frac{j_1+j_2}{2} & m&=\frac{j_1-j_2}{2}\\
  &\hat{\mathcal{B}}^{\frac13 \frac16}_{m,0,j_2} & \D&=\frac{j_2}{2} & m&=\frac{-j_2}{2} & &j_1=0\\
  &\hat{\mathcal{B}}^{\frac16 \frac13}_{m,j_1,0} & \D&=\frac{j_1}{2} & m&=\frac{j_1}{2} & &j_2=0
\end{align}

We conclude by considering the recombination of short multiplets into long ones at the unitarity bound. For $m< \frac{j_1-j_2}{2}$ the unitarity bound is for $\D=\frac13(2j_1+j_2-m)$ and one can verify that
\begin{equation}
 \mathcal{A}^{-\frac13 m+\frac23 j_1+\frac13 j_2}_{m,j_1,j_2}=\mathcal{B}^\frac16_{m,j_1,j_2} \oplus \mathcal{B}^{\frac16}_{m+\frac12,j_1+1,j_2}
\end{equation}
Equivalently, for $m> \frac{j_1-j_2}{2}$ one has
\begin{equation}
 \mathcal{A}^{\frac13 m+\frac13 j_1+\frac23 j_2}_{m,j_1,j_2}=\bar{\mathcal{B}}^\frac16_{m,j_1,j_2} \oplus \bar{\mathcal{B}}^\frac16_{m-\frac12,j_1,j_2+1}
\end{equation}
For the particular case $m=\frac{j_1-j_2}{2}$ we have
\begin{equation}
 \mathcal{A}^{j_1+j_2}_{\frac{j_1-j_2}{2},j_1,j_2}=\hat{\mathcal{B}}^{\frac16\frac16}_{\frac{j_1-j_2}{2},j_1,j_2} \oplus \hat{\mathcal{B}}^{\frac16\frac16}_{\frac{j_1-j_2}{2}+\frac12,j_1+1,j_2}\oplus \hat{\mathcal{B}}^{\frac16\frac16}_{\frac{j_1-j_2}{2}-\frac12,j_1+1,j_2+1}\oplus \hat{\mathcal{B}}^{\frac16\frac16}_{\frac{j_1-j_2}{2},j_1+1,j_2+1}
\end{equation}

The specific cases of vanishing Dynkin labels have to be considered with particular care. Notice that if $j_1=0$ and $Q^1\ket{\D,m,0,j_2}^{\text{hw}}=0$ then the condition $E_1^{-}\ket{\D,m,0,j_2}^{\text{hw}}=0$ automatically implies that $Q^2\ket{\D,m,0,j_2}^{\text{hw}}=0$. Therefore the multiplet $\mathcal{B}^{\frac16}_{m,0,j_2}$ is equivalent to $\mathcal{B}^{\frac13}_{m,0,j_2}$. A similar argument holds for the conjugate case. Based on this arguments we can list all possible multiplets with vanishing labels as
\begin{align}
 &\{\bar{\mathcal{B}}^{\frac16}_{m,0,j_2},\mathcal{B}^{\frac13}_{m,0,j_2},\hat{\mathcal{B}}^{\frac13 \frac16}_{m,0,j_2}\}  & j_1&=0 & j_2&>0 \\
  &\{\mathcal{B}^{\frac16}_{m,j_1,0},\bar{\mathcal{B}}^{\frac13}_{m,j_1,0},\hat{\mathcal{B}}^{\frac16 \frac13}_{m,j_1,0}\}  & j_1&>0 & j_2&=0 \\
  &\{\mathcal{B}^{\frac12}_{m,0,0},\bar{\mathcal{B}}^{\frac12}_{m,0,0}\}  & j_1&=0 & j_2&=0 
\end{align}
For each of these cases the long multiplet at the unitarity bound can be expressed in terms of the short ones. The detailed decompositions are shown in table \ref{longmultdec}.

\begin{table}[htbp]
\begin{center}
\def\arraystretch{1.5}
\begin{tabular}{|c|c|c|}
\hline 
	   	& $m<-\frac{j_2}{2}$  & $\mathcal{A}^{\frac13 (j_2-m)}_{m,0,j_2}=\mathcal{B}^{\frac13}_{m,0,j_2}\oplus \mathcal{B}^{\frac16}_{m+\frac12,1,j_2}$  \\
$j_1=0$		& $m>-\frac{j_2}{2}$ & $\mathcal{A}^{\frac13 (2j_2+m)}_{m,0,j_2}=\bar{\mathcal{B}}^{\frac16}_{m,0,j_2}\oplus \bar{\mathcal{B}}^{\frac16}_{m-\frac12,0,j_2+1}$\\
		& $m=-\frac{j_2}{2}$ & $\mathcal{A}^{\frac{j_2}{2}}_{-\frac{j_2}{2},0,j_2}=\hat{\mathcal{B}}^{\frac13\frac16}_{-\frac{j_2}{2},0,j_2}\oplus \hat{\mathcal{B}}^{\frac13\frac16}_{-\frac{j_2+1}{2},0,j_2+1}\oplus \hat{\mathcal{B}}^{\frac16\frac16}_{\frac{1-j_2}{2},1,j_2}\oplus \hat{\mathcal{B}}^{\frac16\frac16}_{-\frac{j_2}{2},1,j_2+1}$\\[1ex]
\hline 
	\To   	& $m<\frac{j_1}{2}$  & $\mathcal{A}^{\frac13 (2j_1-m)}_{m,j_1,0}=\mathcal{B}^{\frac16}_{m,j_1,0}\oplus \mathcal{B}^{\frac16}_{m+\frac12,j_1+1,0}$\\
$j_2=0$		& $m>\frac{j_1}{2}$ & $\mathcal{A}^{\frac13 (j_1+m)}_{m,j_1,0}=\bar{\mathcal{B}}^{\frac13}_{m,j_1,0}\oplus \bar{\mathcal{B}}^{\frac16}_{m-\frac12,j_1,1}$\\
		& $m=\frac{j_1}{2}$ & $\mathcal{A}^{\frac{j_1}{2}}_{\frac{j_1}{2},j_1,0}=\hat{\mathcal{B}}^{\frac16\frac13}_{\frac{j_1}{2},j_1,0}\oplus \hat{\mathcal{B}}^{\frac16\frac13}_{\frac{j_1+1}{2},j_1+1,0}\oplus \hat{\mathcal{B}}^{\frac16\frac16}_{\frac{j_1-1}{2},j_1,1}\oplus \hat{\mathcal{B}}^{\frac16\frac16}_{\frac{j_1}{2},j_1+1,1}$\\[1ex]
\hline 
\multirow{ 2}{*}{$j_1=j_2=0$}		&$m<0$  & $\mathcal{A}^{-\frac{m}{3}}_{m,0,0}=\mathcal{B}^{\frac12}_{m,0,0}\oplus \mathcal{B}^{\frac16}_{m+\frac12,1,0}$\\
			&$m>0$  & $\mathcal{A}^{\frac{m}{3}}_{m,0,0}=\bar{\mathcal{B}}^{\frac12}_{m,0,0}\oplus \bar{\mathcal{B}}^{\frac16}_{m-\frac12,0,1}$\\[1ex]
\hline
\end{tabular}
\end{center}
\caption{Decomposition of long multiplets into short ones for the case of some vanishing Dynkin labels.}\label{longmultdec}
\end{table}

\subsection{Particular cases: the displacement multiplets}\label{dispmult}
The displacement operator is characterized by quantum numbers $[2,\pm 3,0,0]$. We then look for all the possible short multiplets containing one state with those quantum numbers and with an available highest weight operator. We consider just the case $[2,3,0,0]$ since the negative one can be obtained by simply conjugating the multiplets. In principle the condition of containing one operator with the correct quantum numbers is not very constraining, but in this case, given that the labels are quite small the number of cases is limited. 
We find four of them. 

\subsubsection*{$\bar{\mathcal{B}}^{\frac12}_{\frac32,0,0}$}\label{disp12}
The first multiplet is 1/2 BPS and it is given by.
\begin{center}
\begin{tikzpicture}
 \draw[->]  (-5,5)--(-4.7,4.7);
  \node[above] at (-5.5,5) {$[\frac12,\frac32,0,0]$};
 \draw[->]  (-4,4)--(-3.7,3.7);
  \node[above] at (-4.5,4) {$[1,2,1,0]$};
\draw[->]  (-3,3)--(-2.7,2.7);
  \node[above] at (-3.5,3) {$[\frac32,\frac52,0,1]$};
  \node[above] at (-2.5,2) {$[2,3,0,0]$};
  \end{tikzpicture}
\end{center}
where right arrows indicate the action of a supercharge $Q^a$. Since the only way to create a state of dimension $\frac12$ is with a bosonic field this state is associated to a fermionic component of the displacement operator. In particular the highest weight with the correct quantum numbers is $O=Z$.

\subsubsection*{$\bar{\mathcal{B}}^{\frac13}_{2,1,0}$}
The second multiplet we find is 1/3 BPS. The highest weight has quantum numbers $[1,2,1,0]$ and the descendants are
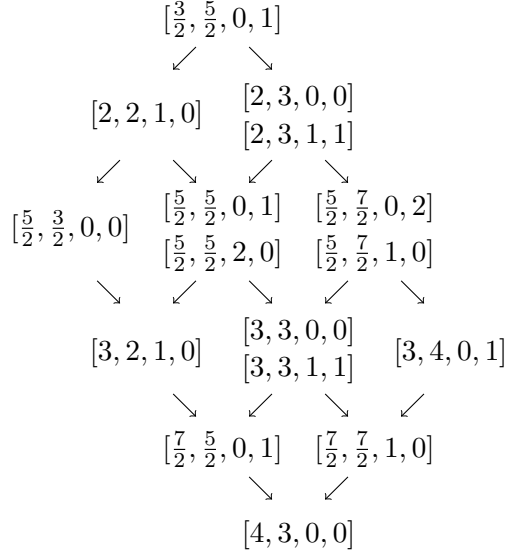
\begin{center}
\begin{tikzpicture}
  \node[above] at (0,5) {$[1,2,1,0]$};
  \draw[->]  (0.35,5)--(0.65,4.7);
  \node[above] at (1,4) {$[\frac32,\frac52,2,0]$};
  \node[above] at (1,3.4) {$[\frac32,\frac52,0,1]$};
  \draw[->]  (-0.35,5)--(-0.65,4.7);
  \node at (-1,4.1) {$[\frac32,\frac32,0,0]$};
  \draw[->]  (1.35,3.4)--(1.65,3.1);
  \draw[->]  (0.65,3.4)--(0.35,3.1);
  \draw[->]  (-0.65,3.4)--(-0.35,3.1);
  \node[above] at (2,2.4) {$[2,3,0,0]$};
  \node[above] at (2,1.8) {$[2,3,1,1]$};
  \node at (0,2.5) {$[2,2,1,0]$};
  \draw[->]  (2.35,1.8)--(2.65,1.5);
  \draw[->]  (1.65,1.8)--(1.35,1.5);
  \draw[->]  (0.35,1.8)--(0.65,1.5);
  \node[above] at (1,0.8) {$[\frac52,\frac52,0,1]$};
  \node[above] at (3,0.8) {$[\frac52,\frac72,1,0]$};
  \draw[->]  (2.65,0.8)--(2.35,0.5);
  \draw[->]  (1.35,0.8)--(1.65,0.5);
  \node[above] at (2,-0.1) {$[3,3,0,0]$};
  \end{tikzpicture}
\end{center}
In this case we may have both bosonic and fermionic states with the correct highest weight quantum numbers. Nevertheless it turns out that the fermionic operator with the correct quantum numbers, i.e. $\bar\chi_+^1$, is a descendant of the short multiplet $\bar{\mathcal{B}}^{\frac12}_{\frac32,0,0}$. Therefore the highest weight has to be bosonic and it is given explicitly by $O=Z\bar Y^1$.

\subsubsection*{$\bar{\mathcal{B}}^{\frac13}_{\frac72,1,0}$}
The third multiplet is also 1/3 BPS. The highest weight has quantum numbers $[\frac32,\frac72,1,0]$ and reads
\begin{center}
\begin{tikzpicture}
  \node[above] at (0,5) {$[\frac32,\frac72,1,0]$};
  \draw[->]  (0.35,5)--(0.65,4.7);
  \node[above] at (1,4) {$[2,4,2,0]$};
  \node[above] at (1,3.4) {$[2,4,0,1]$};
  \draw[->]  (-0.35,5)--(-0.65,4.7);
  \node at (-1,4.1) {$[2,3,0,0]$};
  \draw[->]  (1.35,3.4)--(1.65,3.1);
  \draw[->]  (0.65,3.4)--(0.35,3.1);
  \draw[->]  (-0.65,3.4)--(-0.35,3.1);
  \node[above] at (2,2.4) {$[\frac52,\frac92,0,0]$};
  \node[above] at (2,1.8) {$[\frac52,\frac92,1,1]$};
  \node at (0,2.5) {$[\frac52,\frac72,1,0]$};
  \draw[->]  (2.35,1.8)--(2.65,1.5);
  \draw[->]  (1.65,1.8)--(1.35,1.5);
  \draw[->]  (0.35,1.8)--(0.65,1.5);
  \node[above] at (1,0.8) {$[3,4,0,1]$};
  \node[above] at (3,0.8) {$[3,5,1,0]$};
  \draw[->]  (2.65,0.8)--(2.35,0.5);
  \draw[->]  (1.35,0.8)--(1.65,0.5);
  \node[above] at (2,-0.2) {$[\frac72,\frac92,0,0]$};
  \end{tikzpicture}
\end{center}
By looking at \eqref{fieldcharges1}, \eqref{fieldcharges2} and \eqref{fieldcharges3} and by keeping in mind the color structure we realize that we can combine the fundamental fields to build a highest weight with the correct quantum numbers in a single way: $O=Z\bar Y^1 Z$. We will see that this multiplet does not contribute to the displacement operator.

\subsubsection*{$\bar{\mathcal{B}}^{\frac16}_{\frac52,0,1}$}
The last multiplet is  1/6 BPS. The highest weight has quantum numbers $[\frac32,\frac52,0,1]$ and reads
\begin{center}
\begin{tikzpicture}
  \node[above] at (0,4.9) {$[\frac32,\frac52,0,1]$};
  \draw[->]  (0.35,4.9)--(0.65,4.6);
  \node[above] at (1,3.9) {$[2,3,0,0]$};
  \node[above] at (1,3.4) {$[2,3,1,1]$};
  \draw[->]  (-0.35,4.9)--(-0.65,4.6);
  \node at (-1,4) {$[2,2,1,0]$};
  \draw[->]  (1.35,3.4)--(1.65,3.1);
  \draw[->]  (0.65,3.4)--(0.35,3.1);
  \draw[->]  (-0.65,3.4)--(-0.35,3.1);
   \draw[->]  (-1.35,3.4)--(-1.65,3.1);
   \node at (-2,2.5) {$[\frac52,\frac32,0,0]$};
  \node[above] at (0,2.4) {$[\frac52,\frac52,0,1]$};
  \node[above] at (0,1.8) {$[\frac52,\frac52,2,0]$};
 \node[above] at (2,2.4) {$[\frac52,\frac72,0,2]$};
  \node[above] at (2,1.8) {$[\frac52,\frac72,1,0]$};
  \draw[->]  (2.35,1.8)--(2.65,1.5);
  \draw[->]  (1.65,1.8)--(1.35,1.5);
  \draw[->]  (0.35,1.8)--(0.65,1.5);
  \draw[->]  (-0.35,1.8)--(-0.65,1.5);
  \draw[->]  (-1.65,1.8)--(-1.35,1.5);
  \node at (-1,0.85) {$[3,2,1,0]$};
  \node[above] at (1,0.8) {$[3,3,0,0]$};
  \node[above] at (1,0.3) {$[3,3,1,1]$};
  \node at (3,0.85) {$[3,4,0,1]$};
  \draw[->]  (2.65,0.3)--(2.35,0);
  \draw[->]  (1.35,0.3)--(1.65,0);
  \draw[->]  (0.65,0.3)--(0.35,0);
  \draw[->]  (-0.65,0.3)--(-0.35,0);
  \node[above] at (2,-0.8) {$[\frac72,\frac72,1,0]$};
  \node[above] at (0,-0.8) {$[\frac72,\frac52,0,1]$};
  \draw[->]  (1.65,-0.8)--(1.35,-1.1);
  \draw[->]  (0.35,-0.8)--(0.65,-1.1);
  \node[above] at (1,-1.9) {$[4,3,0,0]$};
  \end{tikzpicture}
\end{center}
Highest weight operators for this multiplet are discussed in the next section.

\section{Supersymmetry transformation of the fields}\label{susyonfields}
We now consider supersymmetry transformations of the scalar fields under the preserved supercharges 
\begin{align}
 Q^a Z&=2\bar \chi^a_+     &  \bar Q_a Z&=0 &  Q^a \bar Z&=0  &\bar Q_a \bar Z&=-2\chi^+_a  \\
 Q^a Y_b&=-2 \d^a_b \bar \psi_+ &  \bar Q_a Y_b&=-2 \e_{abc} \bar \chi^c_- & Q^a \bar Y^b&=-2\e^{abc}\chi^-_c & \bar Q_a \bar Y^b&=2\d^b_a\psi^+
\end{align}
and similarly for fermions
\begin{align}
 \bar Q_a \psi^+&=0    & Q^a \psi^+&=-2iD_1 \bar Y^a-\frac{4\pi i}{k}[ \bar Y^a l_B-\hat{l}_B \bar Y^a]  \\
 Q^a \psi^-&=-2D\bar Y^a  &\bar Q_a \psi^-&=-\frac{8\pi i}{k}\e_{abc}\bar Y^b Z \bar Y^c  \\
   \bar Q_a \chi^+_b&=-2\e_{abc}\bar D \bar Y^c & Q^a \chi^+_b&=2i\d^a_b D_1 \bar Z+\frac{8\pi i}{k} [\bar Z\Lambda^a_b-\hat\Lambda^a_b\bar Z]     \\
  Q^a \chi^-_b&=2\d^a_b D \bar Z &\bar Q_a \chi^-_b&=-2i\e_{abc}D_1 \bar Y^c-\frac{4\pi i}{k}\e_{acd}[\bar Y^c  \Theta^d_b-\hat\Theta^d_b \bar Y^c]  \\
   Q^a \bar \psi_+&=0    & \bar Q_a \bar \psi_+&=2iD_1  Y_a+\frac{4\pi i}{k}[ Y_a \hat{l}_B-l_B  Y_a]  \\
 \bar Q_a \bar \psi_-&=-2\bar D Y_a  & Q^a \bar \psi_-&=-\frac{8\pi i}{k}\e^{abc} Y_b \bar Z  Y_c  \\
    Q^a \bar \chi_+^b&=2\e^{abc} D  Y_c & \bar Q_a \bar \chi_+^b&=-2i\d^b_a D_1 Z-\frac{8\pi i}{k} [Z \hat\Lambda_a^b-\Lambda_a^b Z]     \\
  \bar Q_a \bar \chi_-^b&=2\d^b_a \bar D Z & Q^a \bar \chi_-^b&=-2i\e^{abc}D_1 Y_c-\frac{4\pi i}{k}\e^{acd}[  Y_c \hat\Theta_d^b-\Theta_d^b Y_c]
\end{align}
where we used the definitions
\begin{align}
  D&=D_2-i D_3 & \bar D&=D_2+i D_3
  \end{align}
and the entries of the supermatrices
\begin{align}
\begin{pmatrix} \L_a^b  & 0\\
             0 & \hat \L_a^b\end{pmatrix}&=\begin{pmatrix}Y_a\bar Y^b+\frac12\d^b_a l_B   & 0\\
             0 & \bar Y^b Y_a+\frac12\d^b_a \hat{l}_B \end{pmatrix} \\
\begin{pmatrix} \Theta_a^b  & 0\\
             0 & \hat \Theta_a^b\end{pmatrix}&=\begin{pmatrix} Y_a\bar Y^b-\d^b_a (Y_c\bar Y^c+Z \bar Z)  & 0\\
             0 & \bar Y^b Y_a-\d^b_a (\bar Y^c Y_c+\bar Z Z)\end{pmatrix}\\
 \begin{pmatrix} l_B  & 0\\
             0 & \hat{l}_B\end{pmatrix}&=\begin{pmatrix} (Z \bar Z-Y_a \bar Y^a)  & 0\\
             0 &  (\bar Z Z -\bar Y^a Y_a )\end{pmatrix}
\end{align}
Notice that, due to the last identity the bosonic part of the superconnection reads
\begin{align}
 \mathcal{L}_B=\frac{2\pi i}{k} \begin{pmatrix} l_B  & 0\\
             0 & \hat{l}_B\end{pmatrix}
\end{align}
Finally we can list the transformation properties of the gauge fields
\begin{align}
 Q^a A_1&=\frac{4\pi i}{k} ( \bar \psi_+ \bar Y^a-\bar \chi^a_+ \bar Z- \e^{abc} Y_b \chi_c^-) & \bar Q_a A_1&=\frac{4\pi i}{k} (Z\chi_a^+-Y_a\psi^+ + \e_{abc} \bar \chi^b_- \bar Y^c) \\
  Q^a A&=0 &  \bar Q_a A&=\frac{8\pi }{k} ( Y_a \psi^- - Z \chi_a^- + \e_{abc} \bar \chi^b_+ \bar Y^c)\\
  Q^a \bar A&=\frac{8\pi }{k} ( \bar \psi_- \bar Y^a-\bar \chi^a_- \bar Z+\e^{abc} Y_b \chi_c^+) &  \bar Q_a \bar A&=0\\
  Q^a \hat A_1&=\frac{4\pi i}{k} ( \bar Y^a \bar \psi_+ - \bar Z \bar \chi^a_+ - \e^{abc} \chi_c^- Y_b ) & \bar Q_a \hat A_1&=\frac{4\pi i}{k} (\chi_a^+ Z-\psi^+  Y_a + \e_{abc}  \bar Y^c \bar \chi^b_-) \\
  Q^a \hat A&=0 &  \bar Q_a \hat A&=\frac{8\pi }{k} (  \psi^- Y_a- \chi_a^- Z+\e_{abc}  \bar Y^c \bar \chi^b_+)\\
  Q^a \bar {\hat A}&=\frac{8\pi }{k} (  \bar Y^a \bar \psi_- - \bar Z \bar \chi^a_-  +  \e^{abc}  \chi_c^+ Y_b) &  \bar Q_a \bar{\hat A}&=0
\end{align}

To check the closure of these transformations and to use them on local operators it is important to keep in mind the equations of motion.
For the gauge field we are interested in the components \eqref{fieldstrenghtcomps} of the field strenght. In particular we focus on the first one, which respects the equation
\begin{equation}\label{Feqmot}
 \mathcal{F} =\frac{2\pi i}{k} \begin{pmatrix} Z\overleftrightarrow{D}\bar Z+ Y_a \overleftrightarrow{D}\bar Y^a+2\bar\psi_+\psi^- +2\bar\chi_+^a\chi^-_a & 0\\
            0& -\bar Z\overleftrightarrow{D}Z- \bar Y^a \overleftrightarrow{D}Y_a -2\psi^-\bar\psi_+ -2\chi^-_a\bar\chi_+^a
            \end{pmatrix} 
\end{equation}
where the operator $\overleftrightarrow{D}$ has the usual definition $Z\overleftrightarrow{D}\bar Z\equiv ZD\bar Z-DZ\bar Z$.
For the fermions we need the equation
\begin{equation}
 \slashed{D}\psi_J=\frac{2\pi}{k} \left(\bar C^I C_I\psi_J-\psi_J C_I \bar C^I +2 \psi_I C_J \bar C^I-2 \bar C^I C_J \psi_I + 2\e_{ILKJ} \bar C^I \bar \psi^L \bar C^K\right)
 \end{equation}
whose projection yields (we list just the components we needed for our computations)
\begin{align}\label{fermeq}
 D\psi^+ = i D_1 \psi^- &+ \frac{2\pi i}{k}\left( \hat l_B \psi^- - \psi^- l_B +2\bar Y^a Z\chi^-_a -2 \chi^-_a Z \bar Y^a-2\bar Y^a \bar \chi_+^b \bar Y^c \e_{abc}\right)\\
 D \chi _a^+=i D_1\chi ^-_a &+\frac{2 \pi i}{k}  \left(\chi ^-_b \O_a^b- \hat \O_a^b \chi ^-_b-2 \bar{Z} Y_a \psi ^-+2 \psi ^-
   Y_a \bar{Z}+\right)\\
   &+\frac{4 \pi i}{k}\epsilon _{a c d} \left(\bar{Y}^c \bar{\psi }_+ \bar{Y}^d+\bar{Y}^d \bar{\chi }_+^c \bar{Z}-\bar{Z} \bar{\chi }_+^c \bar{Y}^d\right)
   \end{align}
with 
\begin{equation}
 \O_a^b= \Theta_a^b+ \Lambda_a^b-\frac12 \d_a^b l_B
\end{equation}

Given the supersymmetry transformations and the equations of motion we can finally consider the four multiplets given in section \ref{dispmult} and apply the appropriate supercharges to recover the components of the superdisplacement operator. A summary is given in table \ref{explmult} below:
\begin{table}[htbp]
\begin{center}
\def\arraystretch{1.5}
\begin{tabular}{|c|c|c|}
\hline 
Multiplet  &  Highest weight & Displacement candidate\\
\hline 
\To
$\bar{\mathcal{B}}^{\frac12}_{\frac32,0,0}$	& $O=Z$  & $\frac16\e_{abc}Q^aQ^bQ^c O=-8 D\bar\psi_+$\\
$\bar{\mathcal{B}}^{\frac13}_{\frac72,1,0}$	& $Z\bar Y^a Z$ & $\frac13\bar Q_a Z\bar Y^a Z=2 Z\psi^+ Z$\\
$\bar{\mathcal{B}}^{\frac13}_{2,1,0}$		& $Z\bar Y^a$ & $\frac16\e_{abc}Q^aQ^bZ\bar Y^c=4(\frac13 DY_d\bar Y^d-ZD\bar Z-\frac23 \bar \chi_+^a \chi^-_a)$\\
\multirow{ 2}{*}{$\bar{\mathcal{B}}^{\frac16}_{\frac52,0,1}$}     & $O_a=\frac12\e_{abc} \bar Y^b Z \bar Y^c$   & $Q^a O_a= 2\bar Y^a Z\chi^-_a -2 \chi^-_a Z \bar Y^a-\bar Y^a \bar \chi_+^b \bar Y^c \e_{abc}$\\
& $O_a=\e_{abc} \bar \chi_+^b \bar Y^c+Z \chi^-_a $   & $Q^a O_a= 2(2 DY_d\bar Y^d+3ZD\bar Z- \bar \chi_+^a \chi^-_a)$\\[1ex]
\hline 
\end{tabular}
\end{center}
\caption{The possible multiplets containing the displacement operator. We apply the appropriate supercharges (according to section \ref{dispmult}) to the highest weight operator and we project the result on the R-symmetry singlet component.}
\label{explmult}
\end{table}

From this table one immediately notices that the supermultiplet $\bar{\mathcal{B}}^{\frac13}_{\frac72,1,0}$ contains an operator which does not appear in the displacement supermatrix \eqref{DBexpl},\eqref{DFexpl}. On the other hand, the bosoninc operators in \eqref{DBexpl} can be found in the two multiplets $\bar{\mathcal{B}}^{\frac13}_{2,1,0}$ and $\bar{\mathcal{B}}^{\frac16}_{\frac52,0,1}$. In particular, the latter admits two possible highest weight operators (actually more than two, but the others are not interesting here) which contain a component of the displacement supermatrix. Indeed, the first line of the last row of table \ref{explmult} contains a set of operators which mixes with $D\psi^+$ (as one can see form the equation of motion \eqref{fermeq}) and therefore contains the bottom left component of \eqref{DFexpl}. The only component left is the top right component in \eqref{DFexpl} which is the easiest one since it clearly appears in the multiplet 
$\bar{\mathcal{B}}^{\frac12}_{\frac32,0,0}$. One may wonder why the operator $\bar \psi_+ \psi^-$ does not show up in table \ref{explmult}. The reason for that is pretty subtle, since in the contest of operator insertions on a Wilson line a conformal descendant is given, as discussed around \eqref{dandD}, by the action of the covariant derivative \eqref{susyvarL}. Therefore, to actually rebuild a conformal descendant of $\psi^-$ in the equation of motion \eqref{fermeq} one would need to write a supermatrix equation with diagonal entries given exactly by $\bar \psi_+ \psi^-$ and $ \psi^- \bar \psi_+$. This shows that the latter operator mixes with those appearing in \eqref{fermeq} in a very subtle way, proving once more the necessary interplay between $su(1,1|3)$ and $U(N|N)$ representations in this contest.
\bibliographystyle{nb}

\bibliography{biblio}

\end{document}